\documentclass[acmsmall]{acmart}
\AtBeginDocument{%
  }

\usepackage{
  float,
  epsfig,
  wrapfig,
  graphics,
  subcaption
}
\usepackage{xspace}
\usepackage{amsmath,amsfonts}
\usepackage{algorithmic}
\usepackage{graphicx}
\usepackage{subcaption}
\usepackage{textcomp}
\usepackage{hyperref}
\usepackage{xcolor}
\usepackage{txfonts}
\usepackage[most]{tcolorbox}
\definecolor{mygreen}{rgb}{0,0.6,0}
\definecolor{mygray}{rgb}{0.5,0.5,0.5}
\definecolor{myblue}{rgb}{0,0,1}
\definecolor{myred}{rgb}{0.6,0,0}
\usepackage{tikz}
\usepackage{graphics}

\usepackage{todonotes}
\usepackage{listings}
\usepackage{tabularx}
\usepackage[noorphans,vskip=0.0pt,leftmargin=.1pt]{quoting}

\lstdefinestyle{base}{
  basicstyle=\ttfamily\scriptsize, % Use a monospaced font
  numbers=none,         % Line numbers on the left
  frame=single,         % Single frame around the code
  commentstyle=\color{mygreen},  % Style for comments
  keywordstyle=\color{myblue},  % Style for keywords
  stringstyle=\color{myred},     % Style for strings
  breaklines=true,        % Enable line breaking
  breakatwhitespace=true,  % Only break lines at white space if possible
}

\usepackage{cleveref}
\crefformat{section}{\S#2#1#3}
\crefname{figure}{Figure}{Figures}
\crefname{table}{Table}{Tables}
\crefname{theorem}{Theorem}{Theorems}
\crefname{thm}{Theorem}{Theorems}
\crefname{lemma}{Lemma}{Lemmata}
\crefname{equation}{Eqt.}{Eqts.}
\crefformat{Grammar}{Grammar #1}
\crefname{appendix}{Appendix}{Appendices}
\crefname{listing}{Listing}{Listings}
    
\setcopyright{acmlicensed}
\copyrightyear{2024}
\acmYear{2024}
\acmDOI{XXXXXXX.XXXXXXX}

\acmConference[ACSAC '24]{}{December 9-13,
  2024}{Waikiki, HI}

\newif\ifDEBUG
\DEBUGfalse
 
\ifDEBUG

\newcommand{\SR}[1]{\todo[color=lightgray,inline]{Sazzadur says: #1}}
\newcommand{\JC}[1]{\todo[color=lime,inline]{Jesse says: #1}}
\newcommand{\JR}[1]{\todo[color=lime,inline]{Jackson says: #1}}
\newcommand{\SN}[1]{\todo[color=lime,inline]{Sathvik says: #1}}

\else 
\newcommand{\JD}[1]{}
\newcommand{\SR}[1]{}
\newcommand{\JC}[1]{}
\newcommand{\JR}[1]{}
\newcommand{\SN}[1]{}
\fi

\newcommand{\ie}{\textit{i.e.\ }}
\newcommand{\eg}{\textit{e.g.\ }}
\newcommand{\etal}{\textit{et al.\ }}

\newcommand{\heart}{\ensuremath\varheartsuit}

\definecolor{carnivalred}{RGB}{171, 5, 32}
\definecolor{azblue}{RGB}{12, 35, 75}
\definecolor{coolgray}{RGB}{226, 233, 235}
\definecolor{warmgray}{RGB}{244, 237, 229}
\definecolor{lightgray}{rgb}{.9,.9,.9}
\definecolor{darkgray}{rgb}{.4,.4,.4}
\definecolor{purple}{rgb}{0.65, 0.12, 0.82}
\definecolor{solablue}{HTML}{0068AD}
\definecolor{solaorange}{HTML}{FF5D00}
\definecolor{darkcandyapplered}{rgb}{0.64, 0.0, 0.0}

\newcommand{\myquote}[1]{\textit{``#1''}}

\newtcolorbox{SummaryBox}[1]{enhanced,arc=1mm,outer arc=1mm,
  boxrule=0mm,toprule=0mm,bottomrule=0mm,left=1mm,right=1mm,leftrule=2pt,
  titlerule=0mm,toptitle=0mm,bottomtitle=0mm,top=0mm,
colframe=blue!50!black,
colback=blue!5!white,
  coltitle=blue!50!black,
  colbacktitle=yellow!50!white,
  colback=blue!5!white,
  title=#1,
  fonttitle=\bfseries\sffamily\normalsize,fontupper=\normalsize\itshape,
}

\newcommand{\posdiff}[1]{\textcolor{green!50!black}{#1}}

\newcommand{\totalprotestwarecount}[0]{32\xspace}

\acmJournal{PACMSE}
\lstdefinelanguage{JavaScript}{
	keywords={typeof, new, true, false, try, catch, function, return, null, catch, switch, var, if, in, while, do, else, case, break, let, const, throw},
	keywordstyle=\color{blue}\bfseries,
	ndkeywords={class, export, boolean, throw, implements, import, this},
	ndkeywordstyle=\color{darkgray}\bfseries,
	identifierstyle=\color{black},
	sensitive=false,
	comment=[l]{//},
	morecomment=[s]{/*}{*/},
	commentstyle=\color{darkgray}\ttfamily,
	stringstyle=\color{carnivalred}\ttfamily,
	escapeinside={/*\#}{\#*/},	
	morestring=[b]',
	morestring=[b]",
	morestring=[b]`
}

\begin{document}

% \tableofcontents

% \title{A First Look at Protestware} 
\title{On the Nature and Impacts of Protestware}
%\title{Protestware: Needs of Many or Needs of One?}
\title{An Investigation into Protestware}

\author{Tanner Finken}
\orcid{}
\affiliation{%
  \institution{University of Arizona}
  \city{Tucson}
  \country{USA}
}
\email{finkent@arizona.edu}

\author{Jesse Chen}
\orcid{0009-0001-1741-5531}
\affiliation{%
  \institution{University of Arizona}
  \city{Tucson}
  \country{USA}
}
\email{jessechen@arizona.edu}

% \author{Sathvik Reddy Nookala}
% \orcid{}
% \affiliation{%
%   \institution{University of Arizona}
%   \city{Tucson}
% \email{sathvikn@arizona.edu}

% \author{Jackson Roby}
% \orcid{}
% \affiliation{%
%   \institution{University of Arizona}
%   \city{Tucson}
% \email{jacksoncroby@arizona.edu}

\author{Sazzadur Rahaman}
\orcid{0000-0002-1258-6470}
\affiliation{%
  \institution{University of Arizona}
  \city{Tucson}
  \country{USA}
}
\email{sazz@cs.arizona.edu}

\renewcommand{\shortauthors}{Finken et al.}
\begin{abstract}
Protests are public expressions of personal or collective discontent with the current state of affairs. 
Although traditional protests involve in-person events, the ubiquity of computers and software opened up a new avenue for activism: protestware.
% Protestware is a piece of software that is utilized for protest to either spread a message or negatively impact a particular group of people due to some ideology. 
The roots of protestware date back to the early days of computing. However, recent events in the Russo-Ukrainian war has sparked a new wave of protestware. 
% In particular, the controversy that sparked a large number of the protestware is the Russia-Ukraine war, where developers were deliberately denying service to organizations and people in Russia. 
While news and media are heavily reporting on individual protestware as they are discovered, the understanding of such software as a whole is severely limited.
In particular, we do not have a detailed understanding of their characteristics and their impact on the community.
To address this gap, we first collect \totalprotestwarecount samples of protestware.
Then, with these samples, we formulate characteristics of protestware using inductive analysis. 
In addition, we analyze the aftermath of the protestware which has potential to affect the software supply chain in terms of community sentiment and usage.
We report that: 
(1) protestware has three notable characteristics, namely, 
i) the ``nature of inducing protests'' is diverse, 
ii) the ``nature of targeting users'' is discriminatory, and 
iii) the ``nature of transparency'' is not always respected;
(2) disruptive protestware may cause substantial adverse impact on downstream users;
(3) developers of protestware may not shift their beliefs even with pushback; 
(4) the usage of protestware from JavaScript libraries has been seen to generally increase over time.
% We also identify some of the key components that compose protestware using deductive analysis and discuss how one might utilize these components to identify if protestware is present in a given open source program. 
% \JC{Made some minor edits above.}
% Once identified, the next step is studying how to defend or prevent this type of attack on computers utilizing open source software. 
% Specifically, in this paper, we performed the analysis on a dataset of 31 identified cases of protestware. 
% First, basic information was collected for each protestware. 
% Then, we inductively identified characteristics of each based on our coding scheme. 
% Finally, we performed temporal analysis to learn if any patterns exist for when protestware remains and how it affects the trust of the user base.
% Additional analysis was performed to address multiple perspectives such as the main programming language, function, and popularity of these open source repositories. 
% The protestware showed a propensity toward web based applications in both language and functionality. 
% We also revealed that most of the time, the popularity increased across time using GitHub statistics, showing a potential of misplaced trust in open source developers.

\textit{[Content Warning: This paper contains aggressive and derogatory language in the form of examples from GitHub user comments, which some might find unsettling.] }
\end{abstract}


  
\maketitle

\section{Introduction}

% \JC{Some devs do sanctioning or complete removal of their software in act of protest, but this isn't really protestware. What if this study was extended to include any type of protesting in software industry?}
% \JC{Perhaps study how the protests affected the software supply chain with case studies, how many software relies on them, etc?}
%The difference analysis might be done by Sazz or Sathvik if possible
% \JC{RQ: What do developers think of protestware? Can be done via study or looking for comments online. What is the line between accceptable and not-acceptable protestware? Do you think protestware or in-person protests are more xyz?}

Protest is a deeply ingrained form of expression in which individuals voice their dissatisfaction with societal issues, either personal or collective, typically through marches and rallies, hoping to inspire others to join their cause~\cite{covidprotestchina,uapalestinianprotest,venezualaprotest}. With technological advancement, the essence of protest has evolved. Software being the driving force for modern technology, programmers wield a unique form of expression, not through banners or chants, but through the very code they craft. This novel concept, often termed as ``protestware'', represents a fusion of activism and technology, reshaping how we perceive and participate in modern dissent~\cite{protestwareisontherise}.

% \JC{Conflict: This novel concept...protestware is as old as computers.}

% The first relatively known existence of protestware dates back to 2014. 
Protestware is as old as computers, with notable examples dating all the way back to 1964. During the Berkeley Free Speech Movement, students took control of the computers by punching holes in their own cards~\cite{protestwarehistory}.
Fast-forwarding to the 2020s, several recent incidents of turning popular ``benign'' open-source software (OSS) libraries into protestware during the Russian-Ukrainian conflict have raised concerns about the security and trustworthiness of OSS for the upcoming future~\cite{protestwareisontherise,styledcomponentsarticle}. 
% \SR{Add the impact of node-ipc on software supply-chain.} 
% \JC{Added}
Perhaps the most notable, node-ipc, an OSS library with over one million weekly downloads, was altered to delete all files for computers in Russian and Belarus~\cite{protestwareontherise}.
It even affected popular OSS libraries like Vue.js~\cite{nodeipcbleepingcomputer}, which is downloaded over 5 million times every week~\cite{vuejsnpm}.
% \JC{We may want to remove Vue.js from here since it's in a case study. Otherwise it may downplay our case study.} \TF{I think it is fine because we perform separate case studies and allows us to connect all the information.}
This highlights the need for an in-depth understanding of \textit{``the incentives, characteristics, and aftermath of protestware''}, which would help us evaluate both its potential as a form of protest and the risks tied to the erosion of trust within the OSS community.

Although there exists a wide skepticism around protestware because of media attention~\cite{protestwareisontherise,colorsarticle,styledcomponentsarticle,leftpad,blogExample} ---  surprisingly, they did not receive much engagement from the research community. To the best of our knowledge, only two works specifically focused on protestware~\cite{kula2022in, Protestware}. Kula~\etal proposes 3 categories (malignant, benign and developer sanctioning) of protestware by giving a few examples of each without systematically collecting a comprehensive set of protestware~\cite{kula2022in}. Cheong~\etal proposed ethical guidelines for the OSS community for protesting. However, an in-depth scrutiny of protestware characteristics and the aftermath is non-existent.
 
To address this gap, we systematically collect \totalprotestwarecount protestware~(\cref{sec:collection}). 
Then, we preprocessed our data by recording analytic memos and summaries on the resulting dataset~(\cref{sec:dataprep})~\cite{guest2012applied}.
Using said data, we consider the following research questions.

\begin{itemize}

    \item \textbf{RQ1--\cref{sec:char}:} 
    What are the prominent characteristics of protestware? 
    Specifically, 
    (RQ1.1--\cref{sec:natureofinducingprotests}) how are the protests induced in protestware?;
    (RQ1.2--\cref{sec:natureoftargetingusers}) do protestware target all users or only specific groups?; and
    (RQ1.3--\cref{sec:natureoftransparency}) are protestware developers transparent about their protests?
    
    % \item \textbf{RQ1.5:} Why do developers create these protestware?
    % \item \textbf{RQ2:} How do the dynamics (popularity, usage, and sentiment) change after abusing the trust by turning a benign library into protestware?
    \item \textbf{RQ2--\cref{sec:aftermath}:} 
    How do the dynamics change after a protestware is induced?
    Specifically, we focus on altered software and their effects on 
    (RQ2.1--\cref{sec:supplychaineffects}) the supply chain, 
    (RQ2.2--\cref{sec:reactionanalysis}) sentiment, and 
    (RQ2.3--\cref{sec:popularityanalysis}) usage trends.
    
\end{itemize}

For \textbf{RQ1}, we iteratively create and adjust themes for different characteristics found in our protestware dataset~(\cref{sec:coding})~\cite{guest2012applied}. 
In RQ1.1, 
% where we study the different ways protests are induced in protestware, 
we created a taxonomy~(\cref{fig:impl:taxonomy}) containing 13 codes with 4 themes.
The largest theme, altered software, contains 15/\totalprotestwarecount items and the most frequent code, ideology promotion, contains 6 items.
Next, for RQ1.2, we found 18/\totalprotestwarecount protestware target all users, with the remaining 14 targetting specific users.
Finally, for RQ1.3, we found 15/\totalprotestwarecount protestware are not transparent, which may cause problems for the developers and undermine their trust.
% construct a taxonomy, as shown in~\cref{fig:impl:taxonomy}, by following 

In \textbf{RQ2}, we first conduct a retrospective observational study to understand the degree to which protestware affected the supply chain of real-world software by using a comprehensive collection of news articles, blogs, and community comments. We report that \texttt{left-pad} caused hundreds of dependency failures per minute and \texttt{node-ipc} caused deletion of Russian users' computer data. Then, we look at the sentiment of the OSS community towards protestware. We found a mixture of positive (3) and negative (8) sentiments regarding the protestware. Despite negative feedback, 5 resisted conforming to these changes. Finally, we looked at the usage trends to understand how the post-protestware ``trust dynamics''. We noticed that the number of dependency counts was generally increasing, even for the ones with active protestware components, which was an unexpected finding.

% and NPM download counts to study the aftermath of the protestware that alters software and documentation.
% For RQ2.1, These two examples show the possible worst-case scenarios for protestware. Next, regarding RQ2.2, we found a mixture of positive(3) and negative(8) sentiments regarding the protestware.
% Despite negative feedback 5 resisted conforming to these changes.
% \JC{@Tanner, can you check how many is this X and include the protestware names in a comment. Can you share the sentiment analysis spreadsheet with me too.}.\TF{It is in the same spreadsheet under Aftermath data sheet}
% Lastly, in RQ2.3, we saw that usage counts were generally increasing. %\JC{Verify this.}\TF{Yes, this is true, because all of them had high dependents and about half had higher download counts}
% \JC{I find my writing above for RQ1 and RQ2 bland, may need to revisit later.} \JC{Are we repeating ourselves too much with the prose above and the summary below?} \TF{I'm not sure, if a summary is done this early typically for security papers then its fine. And the contributions below look good.}
In summary, our core contributions are:
% \SR{Add more details in the summary -- too short.}
%\JC{PTAL. Check if it contains too much overlap with RQ1 and RQ2 prose above.}
\begin{itemize}
    \item \textbf{Dataset}: We collected the first known comprehensive dataset, containing \totalprotestwarecount protestware. This dataset can be a starting point for future works on protestware.
    \item \textbf{Characteristics}: We propose three characteristics of protestware, including one where we create a taxonomy on the ways protests are induced. The other two characteristics describe how the protestware target the users and if they're transparent about their protestware.
    \item \textbf{Aftermath}: We studied the effects of protestware on the supply chain, sentiment, and usage trends. We highlight 2 protestware that had significant impact on their downstream users.
\end{itemize}

\section{Background and Definitions}

\subsection{Protestware}
We define protestware as any software artifacts used for protesting. Protestware has been defined in previous work~\cite{Protestware} as a form of supply chain attack in which the developers of some open-source project deliberately modify their project to cause some impact to all or some utilizing it. Unlike previous work, we extended the definition to include self-sabotage, developer sanctions, and dedicated software that enables one to protest. This definition was expanded because we aim to look at all types of protests through software to comprehensively understand the associated risks and potentials.

%  Typically, this form of attack involves the developer relaying their ideology or opinions to users of their product, though the method and target of this action vary between instances. Some of the most prominent protestware involve anti-war protests, which target Russian users~\cite{nodeipcCodeOriginal}.

% Two other taxonomies exist for the purpose of only categorizing this attack behavior\cite{kula2022in}. In another case of protestware called self-sabotage, where a developer removes all code from a given platform, the protest is seen through the crashing of any code that depends on the removed code, and the developer knows it will affect other projects, thus giving some proof of protest. Developer sanctions are when a company refuses to sell software to a particular group of people, typically in protest to some event or beliefs. The final case of protestware is dedicated software which are code bases which enable protest to be performed from the usage of their software. This kind of software is typically open about the kind of protest they enable and can be implemented in numerous ways, like a library that needs to be imported, an app that helps you find in-person protests, and many others.  

\subsection{Protestware vs Supply Chain Attacks}
A software supply chain refers to the collection of dependencies a software system relies on to operate~\cite{SupplyChainAttack}. A supply chain attack occurs when an attacker compromises this dependency network to control the operations of a target software~\cite{SCA}. Typically, supply chain attacks rely on abusing the trust of their downstream users, who use a software dependency controlled by the attacker~\cite{duan2020towards}.

According to our definition of protestware, not all protestware relies on compromising supply chains. Even the ones that rely on it -- differ from typical supply chain attacks studied in the literature~\cite{duan2020towards}. This is because protestware typically rely on abusing an \textit{established trust} by altering existing popular libraries and, thus, do not require explicit infiltration into the supply chain, whereas other supply chain attacks~\cite{duan2020towards} rely on tricking users to be included in the supply chain. Thus, by design, defending against protestware-induced supply chain attacks is challenging as it would require continuously monitoring the actions and intent of trusted developers. Traditional security measures are not designed for this, i.e., to detect a shift in the motives of trusted entities in the OSS community.

\section{Protestware Collection and Preparation}
\label{sec:collection}
% Updated 5/23/2024
% \JC{@Tanner, can you verify that all details here are correct. Delete this comment if so. Leave a comment where details are incorrect.}
\subsection{Data Collection}
\textbf{Method.} A comprehensive list of protestware is a prerequisite for our study. Since, no universally recognized list of protestware exists---to create this list, we resort to Internet search. Specifically, we search on Google and Bing using the terms: 
``\texttt{protestware}'',
``\texttt{protestware~list}'',
``\texttt{protestware~examples}''
~\footnote{Data collected on May 14, 2024.}.
We manually review the web pages in the first five pages of results, totaling 77 unique web pages.
Example web pages that presented protestware includes GitHub repos~\cite{dataset}, news articles~\cite{protestwareisontherise}, Q\&A forums~\cite{qnaExample}, and blogs~\cite{blogExample}.
We also searched ``protest'' directly in GitHub, since it is the source of many open-source projects.
For each protestware listed in the webpages, we only include it in our corpus if they fit within our definition of protestware.
To validate each \textit{candidate} protestware, we find the original proof of intent to protest.
Proof of intent to protest may be exhibited through a commit history~\cite{sweetalertcommit}, or blog post~\cite{mongodb} by the protestware authors.
If the corresponding source is no longer available, we use a snapshot from a prior date using Wayback Machine~\cite{Wayback}. 
% \JC{@Tanner, do we know which or how many were like this?}
If the corresponding source does not have any associated code, we refer to the original article and any other associated articles on the web as needed. 
One author collects the protestware, and another author asynchronously validates them.
For any disagreements, the two authors discuss until reaching an agreement.
% \JC{@Tanner, did we have any disagreements, if so how many?} \TF{IDK we didn't track this, we just discussed if we thought it was protestware and why}
Collecting data via internet search has been used as a methodology for research in prior computing works~\cite{chen2024contents, DBLP:conf/uss/StephensonANH023, anandayuvaraj2022reflecting}.

\textbf{Results.} 
We found 295 results with 76 unique links, from which we identified and collected a total of \totalprotestwarecount samples of protestware. From the direct GitHub search, we found $10$ out of $41$ repositories that match our definition; other projects were simply data storage repositories or unimplemented projects. For proof of intent to protest, there were a total of three instances in our dataset where the corresponding source is no longer available, forcing us to use prior snapshots~\cite{nodeipccodeoriginal, peacenotwararchive,fakerarchive}. For one case, there was no corresponding code, instead the intent was expressed through a blog post~\cite{mongodb}.

\subsection{Data Preparation}
\label{sec:dataprep}
Qualitative studies require data to be preprocessed prior to analysis~\cite{miles1994qualitative}.
First, we review the protestware by referring to a combination of relevant data sources, \eg commit messages~\cite{sweetalertcommit}, documentation files (i.e., Readme) of the code repository, or any associated articles from the web~\cite{colorsarticle}~\footnote{The cited examples here are for sweetalert2 and colors.js protestware.}.
In cases where the web page to the source no longer exists (\eg 404 page not found, repository or commit was deleted), we use the WayBack Machine~\cite{Wayback} to fetch a snapshot from an earlier date. Qualitative study methodology traditionally recommends taking analytic memos and summaries to bootstrap the analysis process~\cite{miles1994qualitative}. For this purpose, while reviewing the protestware, we recorded \textit{i)} their software type (library or standalone), \textit{ii)} programming language\footnote{The programming language of each of the protestware was determined using GitHub's provided metrics of language percentage.}, \textit{iii)} primary functionalities and \textit{iv)} summaries of how the protestware does the protesting itself or provides protesting functionality to others.
% \SR{Let's use texttt environment to name protestware.} \TF{sound good}
An example of protestware doing the protesting itself is \texttt{Evolution}, where we recorded \textit{``change background image to anti-Russia''}.
On the other hand, an example of protestware providing functionality to others is \texttt{protestpy}, where we initially recorded ``allows black box over tree''.
These notes result in a logical chain of evidence that facilitates further analysis~\cite{miles1994qualitative}---which were created by two authors. Next, these two authors and an independent author (a total of 3 authors) met to review the produced notes and memos. 
Typos and inconsistencies, if any, are identified and fixed during this meeting. 

\section{Initial Observations}
In this section, we present some basic statistics and the triggers of the resulting protestware dataset prepared in~\cref{sec:dataprep}.

\subsection{Dataset Statistics.}

\begin{wraptable}{R}{0.4\textwidth}
\footnotesize
\vspace{-10pt}
\centering
\caption{Programming languages found in the protestware dataset. Only the primary language is considered.
% \SR{Now that we have space (at least horizontally), can we rework this table to make it less confusing?}
}
\begin{tabular}{lr}
 \toprule
\textbf{Language} & \textbf{Count per language} \\
 \toprule
 JavaSript & 12 \\
\hline
 TypeScript & 5 \\
\hline
 C, PHP & 3 \\
\hline
C++, HTML, Python & 2 \\
\hline
Dart, HCL & 1 \\
% \hline
%  & \\
% \hline
%  & \\
% \hline
%  & \\
% \hline
%  & \\
% \hline
%  & \\
% \hline
%  & \\
% \hline

\toprule
\end{tabular}
\label{fig:language}
\end{wraptable}

Here, we briefly present the analysis results of protestware based on their \textit{i)} their software type (library or standalone), \textit{ii)} programming language, \textit{iii)} primary functionalities, that we recorded during our data preparation phase. In total, our dataset corpus had 15 library modules and 16 standalone software. 9 out of 15 libraries are written in JavaScript language and 5 out of 16 standalone software were written in C/C++. The overall language distribution is shown in \cref{fig:language} with 22 of them being web-based languages like JavaScript, HTML, and PHP. Unsurprisingly, the functionality largely followed a similar pattern, with many being JavaScript libraries to enable different functionalities in web-based applications. For example, many of them relate to pop-up boxes or alerts for the user, such as \texttt{awesome-Prometheus-alerts}, \texttt{SweetAlert2}, and \texttt{yad}, which stands for Yet Another Dialog. Others relate to providing services for user interactions. For example, \texttt{RedisDesktopManager} is a GUI for a desktop application, \texttt{Quake3e} is a game engine, and \texttt{voicybot} is a voice bot for a cloud-based instant messaging service. In~\cref{tab:allProtestware}, we present the software type (Column 2), languages (Column 3), and a brief overview of these functionalities (Column 4) for each of the protestware.

\subsection{Triggers}
\label{sec:triggers}

Here, we define triggers as the event(s) that caused the developers to create the protestware. 
Triggers can be identified by viewing their protesting message or via an article online found from~\cref{sec:collection}.
If none of these reveal the trigger, then we consider the trigger to none or unknown.
We found that protestwares in our collection were mainly triggered due to 
the Russo-Ukrainian war (20/\totalprotestwarecount), one of which originated from the Black Lives Matter movement~\cite{taylor2016blacklivesmatter}.
For instance, the developer of \texttt{styled-components} said \textit{``I had heard that the Russian government was beginning to censor Western news websites and realized that we had a unique opportunity to deliver a concise, informative message via an atypical channel: our npm package installations''}~\cite{protestwareontherise}, altering the library to showing a message to users in a Russian time zone~\cite{styledcomponentssnyk}. 
Similarly, the developer of \texttt{es5-ext} believed that the Russian people \textit{``are not exactly sure what’s going on, and they’re under influence of their propaganda media''} and modified \texttt{es5-ext} to redirect them to accurate sources such as BBC's Tor service~\cite{protestwareontherise}.
The developer of \texttt{event-source-polyfill} claims the same and also recommended BBC's Tor service~\cite{eventsourcecommit}. 

Another cause can be disputes or disagreements with companies (3).
Although a specific trigger event is unknown, we found that the developer of \texttt{faker.js} and \texttt{colors.js} was generally dissatisfied with Fortune 500 companies extensively using free OSS while not giving back to the community~\cite{colorsarticle}. 
Specifically, he said he will be \textit{``no longer going to support Fortune 500s (and other smaller sized companies) with [his] free work''}~\cite{fakerquotearchive}.
He also requested \textit{``a six figure yearly contract or fork the project and have someone else work on it.''}~\cite{fakerquotearchive}.
In response, he denied service in his libraries~\cite{colorsarticle}.
The developer of \texttt{left-pad} was triggered by a patent lawyer asking him to change the name of his project or unpublish it from npm due it sharing the same name as the mobile app known as ``Kik''~\cite{leftpad,leftpadblogarchive}.
In retaliation, he deleted all 273 of his libraries from NPM~\cite{leftpad}.

Other causes (for 8 other protestware) include the
Dakota Access Pipeline~\cite{defunddapl},
a US President~\cite{knox},
fighting for freedom in Iran~\cite{syncmahsa},
COVID-19 lockdown in Israel~\cite{1km-co-il},
licensing issues in Singapore~\cite{freemyinternet}, 
notches in screens~\cite{malnotch}, 
political protests across America~\cite{activistsassemble},
difficulty in organizing protests~\cite{protestory}.
However, more information on the causes are limited for these.
It is unclear what triggered the \texttt{ProtestPy} protestware into existence~\cite{protestpy}.

\begin{table}[htbp]
    \centering
    \scriptsize
    
    \begin{tabular}{|l|c|c|l|c|c|c|c|}
    \hline %Overarching labels
    \textbf{ } & \multicolumn{3}{c|}{Basic Specifications} & \multicolumn{3}{c|}{\textbf{RQ1}} & \textbf{RQ2} \\
\hline
    \textbf{Name (linked)} & \rotatebox[origin=c]{90}{\textbf{ S/W Type } } & \textbf{ Lang. } & \textbf{ Functionality } & \textbf{\begin{tabular}{c}
        Nature of Inducing \\ Protest
    \end{tabular}} & \rotatebox[origin=c]{90}{\textbf{ Specific }} & \rotatebox[origin=c]{90}{\textbf{ Publicized }} & \rotatebox[origin=c]{90}{\textbf{ Active? }} \\
    
    \hline
    \href{https://github.com/left-pad/left-pad}{left-pad }~\cite{leftpad}
    & L & JavaScript & Padding String  & Halting Services & $\circ$ & $\circ$ & $\circ$ \\
    \hline
    
    \href{https://www.mongodb.com/blog/post/mongodb-assistance-ukraine-shut-down-work-russia}{MongoDB }~\cite{mongodb} 
    & - & - & DB Storage Company & Halting Services & $\bullet$ & $\bullet$ & $\bullet$ \\
    \hline \hline
    
    \href{https://github.com/arendst/Tasmota/}{Tasmota }~\cite{tasmotacommit}
    & S & C & Firmware for OTA communication &  Altered Software &  $\bullet$ &  $\circ$ &  $\circ$ \\ 
    \hline
    
    \href{https://github.com/samber/awesome-prometheus-alerts}{awesome-prometheus-alerts }~\cite{awesomeprometheusalertscommit}
    & S & HTML & Alert Rules Management &  Altered Software & $\bullet$ & $\circ$ & $\circ$ \\
    \hline

    \href{https://github.com/RIAEvangelist/node-ipc}{node-ipc }~\cite{node-ipc} 
    & L & JavaScript & Module for Inter Process Comm. & Altered Software & $\bullet$ & $\bullet$ & $\circ$ \\
    \hline
    
    \href{https://github.com/medikoo/es5-ext}{es5-ext }~\cite{esfiveextcommit}
    & L & JavaScript & Extension of ECMAScript & Altered Software & $\bullet$ & $\circ$ & $\bullet$ \\
    \hline
    
    \href{https://github.com/Yaffle/EventSource}{EventSource }~\cite{eventsourcecommit}
    & L & JavaScript & Event Handling Library &  Altered Software & $\bullet$ & $\circ$ & $\circ$ \\
    \hline
    
    \href{https://github.com/evolution-cms/evolution}{Evolution }~\cite{evoCMS}
    & L & PHP & Content Management Framework & Altered Software & $\circ$ & $\circ$ & $\bullet^*$ \\
    \hline
    
    \href{https://github.com/backmeupplz/voicy}{voicybot }~\cite{voicybot}
    & S & TypeScript & Telegram Voice Bot & Altered Software & $\circ$ & $\circ$ & $\bullet$ \\
    \hline
    
    \href{https://github.com/sweetalert2/sweetalert2}{SweetAlert2 }~\cite{sweetalertcommit}
    & L & JavaScript & Alternative JavaScript Popup Alerts & Altered Software & $\bullet$ & $\bullet$ & $\bullet^-$ \\
    \hline
    
    \href{https://github.com/ec-/Quake3e}{Quake3e }~\cite{quake3e} 
    & S & C & Game Engine &  Altered Software & $\bullet$ & $\circ$ & $\bullet$ \\
    \hline
    \href{https://github.com/RedisInsight/RedisDesktopManager/tree/2022}{RedisDesktopManager }~\cite{redisdesktopmanagercommit}
    & S & C++ & Desktop GUI for Data Management & Altered Software & $\bullet$ & $\circ$ & $\bullet$ \\
    \hline
    
    \href{https://github.com/Qalculate/qalculate-gtk}{Qalculate-gtk }~\cite{qalculatecommit} 
    & S & C++ & Desktop Calculator & Altered Software & $\bullet$ & $\circ$ & $\bullet$ \\
    \hline
    
    \href{https://github.com/v1cont/yad/tree/master/po}{yad }~\cite{yad} 
    & L & C & Dialog Boxes through CLI & Altered Software & $\bullet$ & $\circ$ & $\bullet$ \\
    \hline

    \href{https://github.com/Marak/colors.js}{colors.js }~\cite{colorscommit} 
    & L & JavaScript & Color Styling Library & Altered Software & $\circ$ & $\circ$ & $\bullet$ \\
    \hline
    
    \href{https://web.archive.org/web/20220129022735/https://github.com/marak/Faker.js/}{faker.js }~\cite{fakerarchive} 
    & L & JavaScript & Fake Data Generation & Altered Software & $\circ$ & $\circ$ & $\circ$ \\
    \hline
    
    \href{https://www.npmjs.com/package/e2eakarev?activeTab=code}{e2eakarev }~\cite{e2eakarev} 
    & L & JavaScript & NPM Library & Altered Software & $\bullet$ & $\circ$ & $\bullet$ \\
    \hline
    
    \href{https://github.com/styled-components/styled-components}{styled-components }~\cite{styledcomponentscode} 
    & L & TypeScript & Component Styles & Altered Software & $\bullet$ & $\circ$ & $\circ$ \\
    \hline \hline

    \href{https://github.com/pnpm/pnpm/}{pnpm }~\cite{pnpmchange} 
    & S & TypeScript & Package Manager & Altered Documentation & $\circ$ & $\bullet$ & $\circ$ \\
    \hline

    \href{https://github.com/AntonShevchuk/yandex}{yandex }~\cite{yandex} 
    & L & PHP & Yandex-XML PHP Library & Altered Documentation & $\circ$ & $\bullet$ & $\circ^*$ \\
    \hline
    
    \href{https://github.com/terraform-aws-modules}{AWS Terraform Modules }~\cite{terraformcommit} 
    & L & HCL & Collection of Terraform Modules & Altered Documentation & $\circ$ & $\bullet$ & $\bullet$ \\
    \hline 
    \href{https://github.com/iamolegga/nestjs-pino}{Nestjs-pino }~\cite{nestjs-pino} 
    & L & TypeScript & Logging Program & Altered Documentation & $\circ$ & $\bullet$ & $\bullet$ \\
    \hline \hline
    \href{https://web.archive.org/web/20220317095621/https://github.com/RIAEvangelist/peacenotwar}{peacenotwar }~\cite{peacenotwararchive} 
    & L & JavaScript & Protest Library & Dedicated Software & $\bullet$ & $\bullet$ & $\circ$ \\
    \hline

    \href{https://github.com/seanpm2001/MalNotch?tab=readme-ov-file#MalNotch}{MalNotch }~\cite{malnotch} 
    & S & Python & Notch Alteration Software &  Dedicated Software &  $\bullet$ &  $\bullet$ &  $\bullet$ \\ 
    \hline
    
    \href{https://github.com/guytepper/1km.co.il}{1km.co.il }~\cite{1km-co-il} 
    & S & JavaScript & Protest Locator & Dedicated Software & $\circ$ & $\bullet$ & $\bullet$ \\
    \hline
    
    \href{https://github.com/ProtonHackers/Activists-Assemble}{Activists-Assmeble }~\cite{activistsassemble} 
    & S & JavaScript & Protest Updater and Locator & Dedicated Software & $\circ$ & $\bullet$ & $\bullet$ \\
    \hline
    
    \href{https://github.com/montoyamoraga/protestpy}{protestpy }~\cite{protestpy}
    & S & Python & Package for Protesting & Dedicated Software & $\circ$ & $\bullet$ & $\bullet$ \\
    \hline
    
    \href{https://github.com/techieshark/defunddapl}{defunddapl }~\cite{defunddapl} 
    & S & JavaScript & Bank Support App &  Dedicated Software & $\circ$ & $\bullet$ & $\bullet$ \\
    \hline
    \href{https://github.com/IJustWantSomeFreedom/sync-mahsa}{sync-mahsa }~\cite{syncmahsa}
    & S & TypeScript & Song Playing Website & Dedicated Software & $\circ$ & $\bullet$ & $\bullet$ \\
    \hline
    
    \href{https://github.com/lesterchan/freemyinternet}{freemyinternet }~\cite{freemyinternet} 
    & S & PHP & Plugin & Dedicated Software & $\circ$ & $\bullet$ & $\bullet$ \\
    \hline
    
    \href{https://github.com/Tal-Dahann/Protestory}{protestory }~\cite{protestory} 
    & S & Dart & Protest Generation Android App & Dedicated Software & $\circ$ & $\bullet$ & $\bullet$ \\
    \hline
    
    \href{https://github.com/IndivisibleTemplate/BasicTemplate}{IndivisibleTemplate }~\cite{indivisibletemplate} 
    & S & HTML & Political Website Template & Dedicated Software & $\circ$ & $\bullet$ & $\bullet$ \\
    \hline
    \end{tabular}
    \caption{All \totalprotestwarecount instances of protestware from our dataset. Note that a large portion of software uses web-based programming languages, and many altered services are libraries, which can affect the supply chain security.    \label{tab:allProtestware}
    % \JC{Should ``active?'' be in RQ1? it seems like we label it there. Or should ``coding method'' be before and outside of RQ1, perhaps after ``data prep'' and before ``init ob''?}} 
    }
    \medskip
    \textbf{Legend:} {\Large $\bullet$ }Present, {\Large $\circ$} Absent. L=Library, S=Standalone
    % HS=Halting Services, AS=Altered Software, AD=Altered Documentation, DS=Dedicated Software, -=N/A. 
    Additionally, $\bullet^*$ indicates the repository is still active, but has an additional repository without the protestware, and $\circ^*$ means the repository has been archived with protestware and is no longer in use.

    \vspace{-1em}
\end{table} %Give table first, then describe it -TF

\section{RQ1: Characteristics}
\label{sec:char}
% \JC{Quote style inspiration: https://dl.acm.org/doi/pdf/10.1145/3510003.3510111}

In this section, we present the protestware characteristics from three different perspectives, \textbf{i)} ways of inducing protest, \textbf{ii)} nature of targeting users, and \textbf{iii)} nature of transparency. The summary of the finding is presented in Table~\ref{tab:allProtestware}.

% \JC{\cite{miller2022did} mentions 4 author types in \$6. Anyway we can use them in our study?}

\subsection{Coding Methodology} 
\label{sec:coding}

% Here, we present our thematic analysis methodology to curate a set of characteristics. 
Our methodology follows an iterative process, in which themes are created and adjusted throughout the entire analysis phase~\cite{guest2012applied}.
First, we define characteristics as important properties that help distinguish between different types of protestware, though these properties may not be unique to protestware only.
Knowing these definitions, we proceed with qualitatively curating codebooks for different characteristics, by reviewing our entire set of protestware.
This process starts with creating our initial codebook by reviewing our memos, and other artifacts such as source code, commit messages, documentation files, web articles, etc. We gather codes for protestware characteristics by asking, ``What are ways to characterize protestware that explain useful information about the software?''. After that, the identified codes were discussed as a group to determine if we unanimously agreed they were useful or not. 

After creating the initial codebook, through group discussions, we iteratively refined it to finalize the codebook and create the themes. The goal for refinement was to ensure consistency in specifying and generalizing a given concept. In other words, we avoided our codes to be too specific or too generic. An example of generalization is \textit{Conditional DoS} as a characteristic for altered software. Targeted DoS is defined as the insertion of conditional statements to block users of certain demographics. Initially, we used a laundry list of codes to capture different styles of identifying target demographics, i.e., IP, domain extensions, geographic location, or language. Then we realized that different ideologies can look for different attributes, so we generalized it to capture the essence of isolating any group of users based on any attributes, which would generalize to any future use cases, too. 
Once the codebook was finalized, two authors asynchronously labeled each protestware using the resulting codebook.
In total, the codebook contained 4 categories to label for the \totalprotestwarecount protestware, for a grand total of $4\times\totalprotestwarecount$=$128$ labels: nature of (1) inducing protests, (2) targeting users, (3) transparency, and (4) if it's active or not. 
Upon comparing their labels, there were only 5/128
(4.0\%) instances of disagreeing labels among the authors.
For each disagreement, they discussed their reasoning until reaching a mutual agreement.

\subsection{RQ1.1: Nature of Inducing Protests}
\label{charResults}
\label{sec:natureofinducingprotests}

\begin{figure*}[!h]
    \centering
    \includegraphics[width=1.0\textwidth]{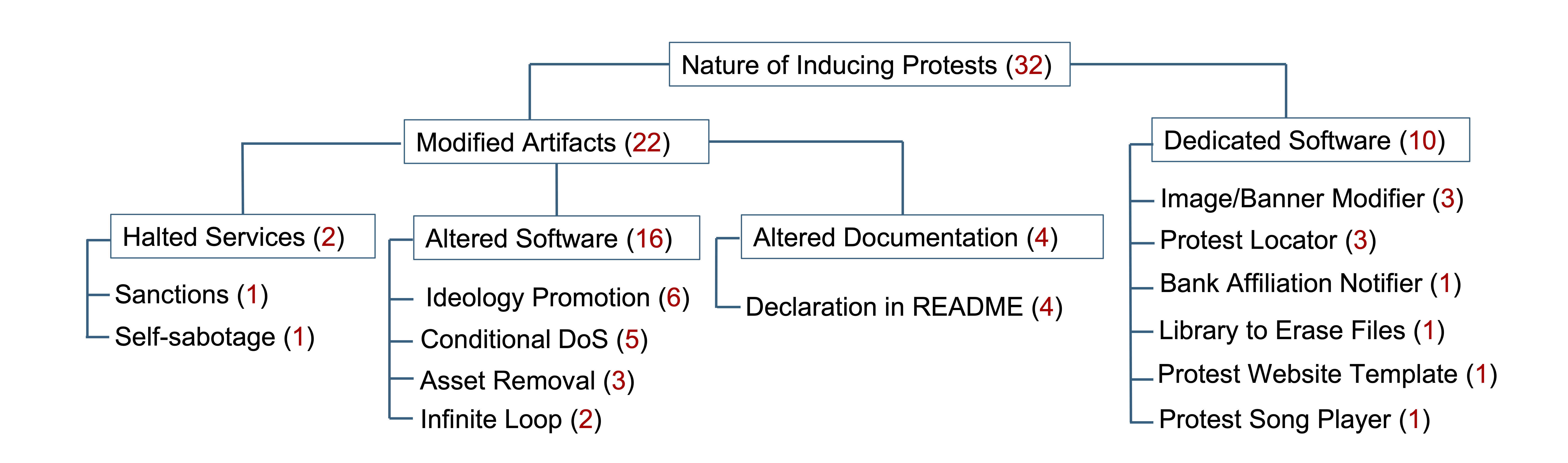}
    \caption{Taxonomy of protestware based on their nature of being induced.
    % \SR{Maybe separate out node-ipc into a separate category, i.e., wiper?} %After Submission
    }
    \label{fig:impl:taxonomy}
\end{figure*}

Our analysis resulted in 13 codes in 4 different themes to capture various ways of enabling protests by the protestware from our corpus (\cref{fig:impl:taxonomy}). First, we found that protestware enables protests in 2 general ways: (1) by modifying artifacts or (2) by developing dedicated artifacts. Artifacts can be modified by either (1a) altering software and/or (1b) documentation or (1c) completely halting service. Next, we discuss our findings in detail. %How does the new numbering look, I think it makes more intuitive sense -TF

% \SR{@Tanner and @Jesse, Add significantly more details under each of the themes -- this is the main meat of the paper, which is too thin right now. For example, we are not discussing the codes for each of the themes, which we must (with examples if needed) -- very important.}

\begin{figure}
  \begin{tabular}{cc}
  \small
    \begin{minipage}[t]{0.48\textwidth}
\lstset{style=base, language=JavaScript}
\begin{lstlisting}[mathescape, caption={Unobfuscated code snippet of Node-IPC. Full unobfuscated code can be found in~\cite{nodeipcCodeUnobf}, and original code can be found in~\cite{nodeipccodeoriginal}.}, label={nodeipcListing}]
setTimeout(function () {
 ...
 if (countryName.includes("russia") 
  || countryName.includes("belarus")) {
    getFiles("./"); getFiles("../"); 
    getFiles("../../"); getFiles("/");
 } ...
}, Math.ceil(Math.random() * 1000));

async function getFiles(...) {
 ... const toDelete = [];
 for (var i=0; i<fileInDir.length; i++){
  ...
  fs.writeFile(combined, "$\heart$", function(){});
  ... } return toDelete; }
\end{lstlisting}
    \end{minipage}
    &
    \begin{minipage}[t]{0.48\textwidth}
\lstset{style=base, language=JavaScript}
\begin{lstlisting}[caption={Code snippet of sweetalert2 showing a custom message ``message.text'' and YouTube video to Russian users.}, label={sweetalertListing}]
// The message will only be shown to Russian users visiting Russian sites
if (navigator.language === 'ru' && 
 location.host
 .match(/\.(ru|su|xn--p1ai)$/)){
  const noWar = document.createElement('div')
  noWar.className = swalClasses['no-war']
  setInnerHtml(
    noWar,
    `<a href="{...${message.youtubeId}}" 
    target="_blank">
    ${message.text}</a>`
    )
  ...
}
\end{lstlisting}
    \end{minipage} \\
  
  \end{tabular}
\end{figure}

% \begin{figure}[h]
% \small
% \lstset{style=base, language=JavaScript}
% \begin{lstlisting}[mathescape, caption={Unobfuscated code snippet of Node-IPC. Full unobfuscated code can be found in~\cite{nodeipcCodeUnobf}, and original code can be found in~\cite{nodeipccodeoriginal}.}, label={nodeipcListing}]
% setTimeout(function () {
%  ...
%  if (countryName.includes("russia") 
%   || countryName.includes("belarus")) {
%     getFiles("./");
%     getFiles("../");
%     getFiles("../../");
%     getFiles("/");
%  }
%  ...
% }, Math.ceil(Math.random() * 1000));

% async function getFiles(path = "", 
%  param2 = "") {
%  ...
%  const toDelete = [];
%  for (var i=0; i<fileInDir.length; i++){
%   ...
%   fs.writeFile(combined, "$\heart$", function(){});
%   ...
%   } return toDelete; }
% \end{lstlisting}
% \end{figure}

% \begin{figure}[h]
% \small
% \lstset{style=base, language=JavaScript}
% \begin{lstlisting}[caption={Code snippet of sweetalert2 showing a custom message ``message.text'' and YouTube video to Russian users.}, label={sweetalertListing}]
% // The message will only be shown to Russian users visiting Russian sites
% if (navigator.language === 'ru' && 
%  location.host.match(/\.(ru|su|xn--p1ai)$/)){
%   const noWar = document.createElement('div')
%   noWar.className = swalClasses['no-war']
%   setInnerHtml(
%     noWar,
%     `<a href="{...${message.youtubeId}}" 
%     target="_blank">
%     ${message.text}</a>`
%     )
%   ...
% }
% \end{lstlisting}
% \end{figure}

\subsubsection{{\bf Altering Software (15).}} 
\label{sec:alteringsw}
Developers may alter existing software to change the services and/or functionality in protest of an issue. Here, we describe in detail the different changes they made.

\textit{Conditional DoS (5).} This is also as the name suggests -- the software denies service based on some condition(s). 
The conditions we observed are all location based.
Perhaps the most notable, \texttt{Node-IPC} overwrites system files with a heart emoji if the user of the software had an IP address located in Russia or Belarus ~\cite{NodeIPC} (code snippet in~\cref{nodeipcListing}).
Other examples also deny services if the user is related to
Russia
~\cite{awesomeprometheusalertscommit, tasmotacommit}.
\texttt{SweetAlert2}, which is a popup box library for JavaScript, disables the expected content in the popup box if the user is a Russian user (\ie navigator.language === ``ru'') visiting Russian sites (\eg .ru, .su)~\cite{sweetalertcommit}. The corresponding code snippet is presented in~\cref{sweetalertListing}. This is also the only observed sample where the changes were implemented via a pull request.
Similarly, \texttt{awesome-prometheus-alerts} removes access to the website for Russian speaking users, directing them to a file called [middle-finger-emoji].md~\cite{awesomeprometheusalertscommit}.
\texttt{Tasmota} also blacklists Russian users not only via language but also location ~\cite{tasmotacommit}.
These types of conditional checks seem to isolate a particular group of users based on factors location, IP, or language checking, with the resulting behavior not being broadly applied to all groups. Therefore, it cannot be an accessibility feature which can perform different functionality depending on some group the user belongs to, typically performed with a \textit{switch} or \textit{if-else} chain.
To demonstrate this, \cref{isofilt} shows a code snippet where code implementing a restriction based on a set of locations around Russia was added for \texttt{es5-ext}~\cite{esfiveextcommit}, showing how no other check was performed to give similar functionality to the groups not represented by this check.

\begin{figure}[!ht]

    \centering
    \includegraphics[width=0.8\linewidth]{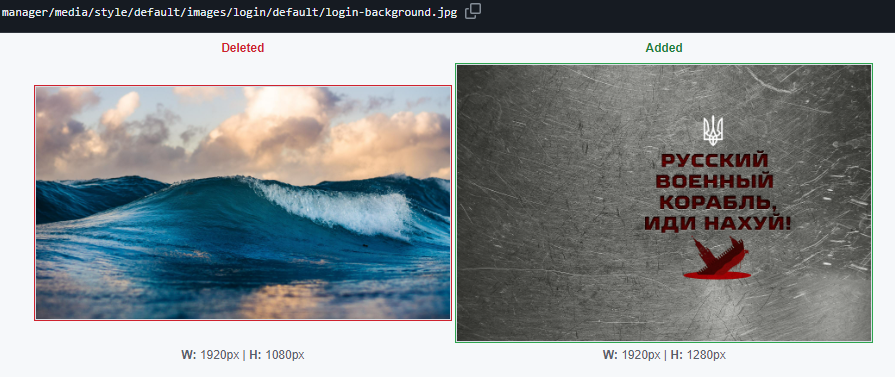}
    \caption{Example of ideology promotion~\cite{evoCMScommit}.}
    \label{evolutioncmsscrnshot}
\end{figure}

\textit{Ideology Promotion (5).} An ideology promotion occurs when when the change to the software promotes some type of belief. 
In \texttt{Evolution} CMS, a background image is changed to a political image with aggressive language against Russia~\cite{evoCMScommit}, as seen in~\cref{evolutioncmsscrnshot}.
\texttt{EventSource}~\cite{eventsourcecommit} and \texttt{es5-ext}~\cite{esfiveextcommit} both print out the same thing: Russian and Ukraine flags and a message in Russian criticizing the Russians invasion on Ukraine and supporting Ukraine.
In \texttt{voicybot}, a string promo text is displayed in both Russian and English: \myquote{Putin and his cronies [kill civilians](https://t.me/verkhovnaradaofukraine) in the war in Ukraine \#stopputin}~\cite{voicybotcommit}.
When trying to install \texttt{styled-components} v5.3.5, a message again criticizing Russia's war on Ukraine is shown~\cite{styledcomponentsarticle}. This message is written in a file named ``\texttt{postinstall.js}'', which the publisher forgot to include in v5.3.4, breaking many builds~\cite{styledcomponentsarticle,styledcomponentsissues}. 
As it can be seen the communication may or may not require the software to be run to be seen and stored in various locations in the software (promotional items, images, console text).
Another data point, although not marked for ideology promotion since it exhibited primarily Targeted DoS, \cref{adcomm} shows how a promotion can be performed with a code snippet in \texttt{awesome-prometheus-alerts} with aggressive language against Russians~\cite{awesomeprometheusalertscommit} in addition to only providing this message toward Russians. 
In addition, this code snippet is saved in a file called [Middle-finger-emoji].md.

Additionally, we would like to note that some instances of sharing the ideology have text in a different language than the primary language of the repository\footnote{Primary language is determined by the language of comments and text files like README.md. Also, the text for spreading the ideology may be a translation of desired message.}. The examples already seen are \cref{evolutioncmsscrnshot} having Russian text in the image, and \cite{voicybotcommit} having text in both English and Russian. 

\begin{figure}[h]
    \centering
    \small 
    \begin{lstlisting}[style=base, 
        language=html, 
        label={adcomm},
        caption={Example of ideology promotion in awesome-prometheus-alerts~\cite{awesomeprometheusalertscommit}.}]
<p>
    [Ukraine flag] Forbidden to Russian people.
    <br>
    <br>
    Please come back as soon as pease returns to Eastern Europe. [Ukraine flag] -
       -[Handshake emoji] [Russian flag]
</p>
    \end{lstlisting}
\end{figure}

\textit{Asset Removal (3)} occurs when the developer deletes anything from the software, \eg image, text, translation support. We observed removal of Russian flags in \texttt{Qalculate}~\cite{qalculatecommit} and removal of Russian translation features[\texttt{Redis},\texttt{yad}]~\cite{redisdesktopmanagercommit, yadcommit}.
% It can be see in \cref{redisnorustrans}\SR{Broken ref}\TF{The figure is in 'results-rq3-old.tex' if we want to include it, otherwise, we can delete this sentence.} that the Russian translation file was removed with a commit message commenting on innocent people dying in Ukraine.  %Do we replace this sentence

\textit{Infinite Loop (2).} In this category, as the name suggests -- an infinite loop was added to the software. Examples are \texttt{colors.js}~\cite{colorscommit} and \texttt{fakers.js}~\cite{colorsarticle}. 
A protestware uses an infinite loop in their code if their goal is to prevent further execution. In turn, this behavior denies the execution of any software relying on it by consuming all it's resources. 
To specify details of \texttt{colors.js} further, we show in~\cref{colorsinfloop} 
% \SR{Broken ref}\TF{Same thing here, I included this because it seems more important but may need to be trimmed to meet page requirement} 
the infinite-loop that was added to freeze the software and show the ASCII image in~\cref{colorsflag} in the supplemental materials~\footnote{A replication package is uploaded with the submission and a link will be provided upon acceptance.\label{fn:appendix}}. 
The modifications to \texttt{colors.js} were to protest large-corporations profiting off of free software without giving back~\cite{colorsnomorefreework}.
\cref{etwoeakarevListing} shows a code-snippet from \texttt{e2eakarev}, where a custom message (PROTEST\_MESSAGE) is printed to Israeli users~\cite{etwoeakarevNPM}. 
Implementation for Python prints the same message~\cite{etwoeakarevPyPI}.

\begin{figure}
  \begin{tabular}{cc}
  \small
    \begin{minipage}[t]{0.48\textwidth}
       \begin{lstlisting}[style=base, 
        language=html, 
        label={colorsinfloop},
        caption={Code snippet of colors.js~\cite{colorscommit}.}]
let am = require('../lib/custom/american');
am();
for (let i = 666; i < Infinity; i++;) {...}
    \end{lstlisting}
    \end{minipage}
    &
    \begin{minipage}[t]{0.48\textwidth}
\begin{lstlisting}[style=base, caption={Code snippet of e2eakarev~\cite{etwoeakarevNPM}.}, label={etwoeakarevListing}]
...
if (userCountryName.includes("israel")) {  
console.log(PROTEST_MESSAGE) } ...
\end{lstlisting}
    \end{minipage} \\
  
  \end{tabular}
\end{figure}

% \begin{figure}[H]
%     \centering
%     \small 
%     \begin{lstlisting}[style=base, 
%         language=html, 
%         label={colorsinfloop},
%         caption={Example of function denial in colors.js.}]
% let am = require('../lib/custom/american');
% am();
% for (let i = 666; i < Infinity; i++;) {...}
%     \end{lstlisting}
%     \vspace{-1em}
% \end{figure}

% \begin{figure}[h]
% \vspace{-1em}
% \small
% \lstset{style=base, language=JavaScript}
% \begin{lstlisting}[caption={Code snippet of e2eakarev showing a custom message (PROTEST\_MESSAGE) to Israeli users~\cite{etwoeakarevNPM}. Implementation for Python prints the same message~\cite{etwoeakarevPyPI}.}, label={etwoeakarevListing}]
% ...
% if (userCountryName.includes("israel")) {  console.log(PROTEST_MESSAGE) }
% ...
% \end{lstlisting}
% \vspace{-1em}
% \end{figure}

\begin{figure}[H]
    \centering
    \small  
    \begin{lstlisting}[style=base, 
        language=Javascript, 
        label={isofilt},
        caption={Example of targeted DoS in es5-ext~\cite{esfiveextcommit}.}]
if ([ "Europe/Moscow", "Asia/Yakutsk", "Asia/Krasnoyarsk", "Europe/Samara", "Asia/Yekaterinburg", "Asia/Irkutsk", "Asia/Anadyr", "Asia/Kamchatka", "Europe/Kaliningrad", "Asia/Vladivostok", "Asia/Magadan", "Asia/Novosibirsk", "Asia/Omsk"
    ].indexOf(new Intl.DateTimeFormat().resolvedOptions().timeZone) === -1) {return;}
    \end{lstlisting}
\end{figure}

\begin{SummaryBox}{Takeaway \cref{sec:alteringsw}: Alterations are done more to deny services in different forms than `solely' for ideology promotions.}
Only 5/15 protestware's main focus is on ideology promotion, whereas the rest focus on alteration to at least partially deny their services to users of certain demographic targets.
\end{SummaryBox}

\subsubsection{\bf Altering Documentation (4).} 
\label{sec:alteringdoc}
Developers may alter documentation to spread protest memos.
If this was the main form of demonstration, we consider the protestware to be altering documentation.

\textit{Declaration in README (4).}
In total, we found 4 altered documentations, all declared in the README file.
An example is \texttt{pnpm}~\cite{pnpmchange}, where the README.md file was edited to raise awareness and collect funding from supporters of Ukraine~(\cref{alteredDoc}--see~\cref{fn:appendix}).  
In \texttt{Terraform}'s README, their terms of users for those from Russia and Belarus state that, by using the software, they agree that Russia has committed certain crimes~\cite{terraformcommit}. A banner also states the same thing about Russia, using the dedicated software called \texttt{StandWithUkraine}~\cite{standwithukraine}.
Lastly, \texttt{nestjs-pino} protested the war by showing an image of children in a bomb shelter in Ukraine and providing donation links~\footnote{A similar message is printed in the console after installation, but the declaration in README is more notable.}~\cite{nestjs-pino}.

% \begin{figure}
% \small
% \lstset{style=base}
% \begin{lstlisting}[mathescape, 
% caption={Added protest message in README.md in pnpm~\cite{pnpmChange}.}, 
% label={alteredDoc}]
% > # UKRAINE NEEDS YOUR HELP NOW!
% >
% > I'm the creator of this project and I'm Ukrainian.
% >
% > **My country, Ukraine, [is being invaded by the Russian Federation, right now](https://www.bbc.com/news/...)**. I've fled Kyiv and now I'm safe with my family in the western part of Ukraine. At least for now.
% > Russia is hitting target all over my country by ballistic missiles.
% >
% > **Please, save me and help to save my country!**
% >
% > Ukrainian National Bank opened [an account to Raise Funds for Ukraine’s Armed Forces](https://bank.gov.ua/en/news/all...):
% ...
% > You can also donate to [charity supporting Ukrainian army](https://savelife.in.ua/en/donate/).
% >
% > **THANK YOU!**
% \end{lstlisting}
% \end{figure}

\begin{SummaryBox}{Takeaway \cref{sec:alteringdoc}: When utilizing documentation - it's for gaining mass support for the cause.}
The only method developers used to gain support when altering the documentation was by editing the README file as it was easily accessible.
\end{SummaryBox}

\subsubsection{\bf Halting Services (2).}
\label{sec:haltingservices}
Software owners may halt their software services by either deleting it completely (\ie self-sabotage) or sanctioning an entity.

\textit{Sanctions (1).} 
An example of sanction is \texttt{MongoDB}'s removal of their software and services from the Russian market due to the Russian war on Ukraine~\cite{mongodb}.
It is not known what software issues in Russia arose because of the sanction.

\textit{Self-Sabotage (1).} 
On the other hand, an example of self-sabotage is where the developer of \texttt{left-pad} removed all 273 of his packages~\footnote{Per our definition, all of the 273 packages would be considered protestware. However, we only include \texttt{left-pad} in our corpus, as it caused the most damage, and the other will have the same characteristics.} from npm in protest of trademark issues~\cite{leftpad}. The deletion caused ample damage but was quickly reversed by NPM, as detailed in~\cref{sec:supplychaineffects}.

% \JC{The following is also in aftermath, effects on the supply chain, \cref{sec:supplychaineffects}. Consider removing from here to focus on the ways they are induced, and not the resulting effects.}
% The deletion of left-pad broke so much software down the supply chain that npm decided to republish the package.
% Regarding the restoration of left-pad, npm CTO said \myquote{This action puts the wider interests of the community of npm users at odds with the wishes of one author; we picked the needs of the many}~\cite{leftpad}.

% \begin{SummaryBox}{Takeaway [\cref{sec:haltingservices}]: The needs of many outweigh the needs of one when dealing with problematic changes.}
% Entities such as companies place the needs of many over the needs of one when it comes to reverting disruptive protest functionalities.
% \end{SummaryBox}

\begin{SummaryBox}{Takeaway \cref{sec:haltingservices}: Protests can be extreme.}
Halting services may cause serious software systems failures, but owners may not care if their belief is sufficiently strong.
\end{SummaryBox}

\subsubsection{\bf Dedicated Software (10).} % Dedicated Software, previously called New Software
\label{sec:dedicatedsw}
Developers may develop a new piece of software specifically to protest an issue or convenience protesters.
They can import these packages instead of writing their own functionalities for protesting.

\textit{Image/Banner Modifier (3).}
\texttt{MalNotch} protests hardware notches on screens of devices by placing NSFW/NSFL images to the area of the screen covered by the notch~\cite{malnotch}. Users of notched devices are then banned depending on the platform.
\texttt{ProtestPy} is a python library used to place black boxes on top of images in protest of anything to the user's desire~\cite{protestpy}.
\texttt{Freemyinternet} places the FreeMyInternet banner automatically to a developer's WordPress website in protest of new licensing by the Media Development Authority of Singapore~\cite{freemyinternet}.

\textit{Protest Locator (3).}
\texttt{1km.co.il} was developed to enable Israeli citizens to protest in their neighborhoods during Israel's second lockdown~\cite{onekm}. It also protests a bill passed by the parliament to prevent protests more than 1km from their homes.
\texttt{Activists-Assemble} finds a protest, checks its safety, and directs the user to its location~\cite{activistsassemble}. It also provides live Twitter updates to the event.
\texttt{Protestory} allows users to search and create protests around the world based on a user's views~\cite{protestory}.

% \SR{Can we group the description of these codes as ``Others''?}
\textit{Bank Affiliation Notifier (1).}
\texttt{Defund DAPL} is an app that lets users determine if their bank is funding the Dakota Access Pipeline~\cite{defunddapl}. So they can protest by changing banks if necessary.

\textit{Library to Erase Files (1).}
\texttt{Peacenotwar} is a library that will create a file called {WITH-LOVE-FROM-AMERICA.txt} on the desktop containing heart emojis ~\cite{peacenotwararchive}. 
It is also used by \texttt{node-ipc}~\cite{nodeipcimportpeacenotwarcommit}.

\textit{Protest Song Player (1).}
\texttt{Sync-mahsa} is an offline website that plays songs about freedom in Iran~\cite{syncmahsa}, in support of the protest triggered by Mahsa Amini's death.

\textit{Protest Website Template (1).}
\texttt{IndivisibleTemplate} is a template for users to create a political protest website swiftly~\cite{indivisibletemplate}.

\begin{SummaryBox}{Takeaway \cref{sec:dedicatedsw}: There is a wide variety of standalone software built for protesting!}
Standalone software enables protest expression through various channels, online or in person, and functions, from deleting files to playing songs, etc.
\end{SummaryBox}

\subsection{RQ1.2: Nature of Targeting Users}
\label{sec:natureoftargetingusers}
% \SR{Add more details here too.}
%Might insert some text before these to indicate the thought process of each of these.
% Next, we wanted a way to distinguish who was being affected by the protestware. The two general types would be whether it affects everyone or only a subset of users.
To understand how different protestware would affect different users, we labeled the protestware under two different codes: everyone (universal) or only a subset (specific).

% \SR{@Tanner to put counts under Universal and Specific ()?}
\textit{Universal (18).}
Protestware targets are considered "universal" if the modifications made to the original open source project impact all users of that project. 
The user's information is disregarded entirely when determining if the protestware behavior should be active. 
\texttt{Evolution CMS}~\cite{evoCMS} and \texttt{voicybot}~\cite{voicybot} with universal ideology promotion are examples of protestware with a universal way of affecting users. 
Typically, a universal behavior seen was displaying a message to all users, although a few cases existed to eliminate functionality for all.

\textit{Specific (14).}
A protestware is labeled as ``specific'' if the modifications made to the original open source project target a particular subset of users based on some pre-selected factor, such as nationality, affiliation, geographical location, etc.
These directly violate the anti-discrimination clauses of the Open Source License ~\cite{OSI}. An example of this is \texttt{Sweetalert2}~\cite{sweetalertcommit} where people in Russia visiting Russian sites will be shown a ``stop war'' message. In this case, the developer only wants certain behavior for their target audience and this makes it easier to negatively impact only those users by altering functionality, although simply printing a message can also be done. 

\begin{SummaryBox}{Takeaway \cref{sec:natureoftargetingusers}: A significant portion of Protestware are discriminatory -- affects users with particular demographic.}
We find that 14 protestware had behavior only applied to a specific set of people, thus discriminating against certain people who use their software. 
\end{SummaryBox}

\subsection{RQ1.3: Nature of Transparency} 
\label{sec:natureoftransparency}

From the perspective of protestware being transparent, we labeled them under ``publicized'' or ``hidden''.
The part we looked at was the description of the software, typically in a README.md file, and if the protest behavior was documented in that description.

\textit{Publicized (17).}
Protestware are ``publicized'' if the developer makes a noticeable effort to announce their software as protestware such that users do not need to dig for it.
Typically, this effort can be shown via a message in a README file to indicate alterations. 
An example of this open intent to protest can be seen in \texttt{MalNotch}~\cite{malnotch} where the developer explicitly describes the protesting behavior that running the software will do on devices with notches. 

\textit{Hidden (15).}
A protestware is ``hidden'' if the modifications made to the original open-source project are not publicly announced by the developer to the user base.
To identify it, a user would either have to spend time scrutinizing a commit message or code or run the software itself. An example of this behavior was seen in a project titled \texttt{yad}~\cite{yad}, where Russian translation was removed and this behavior was not reflected in the README.md file. Another example is \texttt{es5-ext}~\cite{esfiveextcommit}, which shows protest messages in Russian time zones but does not declare this behavior.

\begin{SummaryBox}{Takeaway \cref{sec:natureoftransparency}: Transparency is expected but not always respected!}
The transparency of a repository helps to gain trust with the user base to help understand behavior in the software instead of being surprised after running the software.
\end{SummaryBox}

\subsection{Correlations in Different Natures}
\label{sec:correlateNatures}

\begin{figure}[h]
% \begin{wrapfigure}{R}{0.6\textwidth} 
    \centering
    \includegraphics[width=0.5\linewidth]{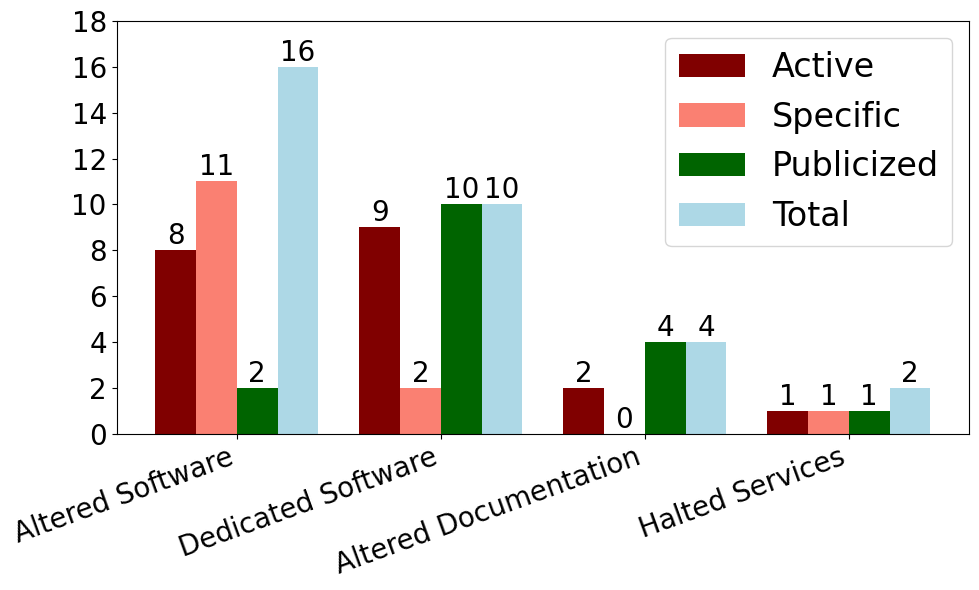} %Figures/breakdownChart.eps} %May need to change format to .eps
    \vspace{-10pt}
    \caption{ Grouped bar chart of protestware breakdowns by our characteristics.
    \JC{Find a better spot for this after reviews.}
    % Initially separated by nature of inducing protest and then shown the values of 1) How many are still active 2) How many affected specific users 3) How many publicized the changes they made and 4) The total amount in each category. 
    % \JC{Put this in wrap fig after finalizing prose. Wrapfig bug happening again.}
    }
    \label{fig:totalBar}
% \end{wrapfigure}
\end{figure}

To better understand how different natures would correlate, we examined the ``nature of transparency'', the ``nature of targeting users'' and ``the operational status'' (active or not), with different ``natures of inducing protest''. Figure ~\ref{fig:totalBar} shows the result. We found that 12 (out of 16) altered software targeted specific users. Only 2 of the 12 targeting specific users had disclaimers about it, and these were the only 2 (out of 16) of the altered software categories that were transparent about it. Understandably, all the dedicated software for the protest was publicized through their documentation, so people knew about their protest behavior, and only one was no longer active (\texttt{peacenotwar}). Altered documentation was also transparent because they changed the documentation of their software.

\begin{SummaryBox}{Takeaway \cref{sec:correlateNatures}: Targeting users for protest is common but not typically publicized.}
Although providing a disclaimer that the software discriminated against a group of users was seen in a few (2) protestware, this was not a common occurrence. Most of the protestware that targeted users were hidden (10 out of 12), waiting for the targeted group to be affected.
\end{SummaryBox}
% \SR{@Tanner, Add a takeaway message?}
% \input{results-rq3}
\section{RQ2: Aftermath Study}

\label{sec:aftermath}
%\SR{Would it make more sense to just focus on the ones from the ``altered software'' here? We can simply provide one paragraph on the ``Altered documentation'' category to show if their popularity increased or not, but that's it.}
%\JC{We need to the entire ``modified artifacts'' since left-pad is in it. We can say what we included in each section.  \eg``In this section, we consider all modified artifacts...''}\SR{Makes sense.}
In this section, we first conduct a retrospective study of protestware's effect on the software supply chain based on the news reported online (\cref{sec:supplychaineffects}).
To understand the consequences and communities' reactions to protestware created by modifying artifacts (21 in total)~\footnote{MongoDB was omitted because it is closed-source.}, we looked at the indicators: i) community reactions (\cref{sec:reactionanalysis}) and ii) usage trends (\cref{sec:popularityanalysis}). We chose these indicators because they help to show the amount of trust people give to these protestware after their modification. 
% \JC{Find some similar prior work for inspiration.}
% \JC{Find comments from reddit using the API.}
% \JC{\cite{miller2022did} mentions 4 author types in \$6. Anyway we can use them in our study?}

\subsection{Impacts on Supply Chain}
\label{sec:supplychaineffects}

Of the \totalprotestwarecount protestware we studied, 8 (5 conditional DoS, 2 infinite loops, 1 self-sabotage) can potentially cause serious problems to any downstream software components. Specifically, we found that 4/5 protestware with conditional DoS would simply not run if the users were Russian~\cite{awesomeprometheusalertscommit,quake3e,sweetalertcommit, tasmotacommit}, while the fifth one is more severe, completely deleting Russian users' computer files (\texttt{node-ipc}~\cite{nodeipcbleepingcomputer}).
The 2 infinite-loops act like DoS's as well~\cite{fakerarchive,colorscommit}. These findings naturally lead to the following research question: ``\textit{To what extent has the protestware contributed to disruptions in the supply chain of real-world software components?}'' Next, we discuss the methodology we designed to answer this question qualitatively and the study's findings.

% To understand to what extent this protestware caused supply chain failure, we conducted a retrospective observational study based on the news online. 
% , news article only briefly mentioned their changes~\cite{colorsarticle} and estimated their effects to be in the range of thousands of developers~\cite{fakersnyk}.
% For instance, \texttt{faker.js} is used to create fake test data.
% While this would be an inconvenience for developers, it would not necessarily affect the end-product.
% In the worst case, users of DoSed libraries can just downgrade versions.
% However, for \texttt{node-ipc}, even though you can still downgrade versions, the deletion of all computer files is generally not reversable. Finally, the self-sabotaged library \texttt{left-pad} saw large number of failures in dependents~\cite{leftpadnpmblog}, but downgrading versions is not possible.

% and only 2 of these had detailed articles.
% Next, we present case studies for the 2 protestware: \texttt{left-pad} and \texttt{node-ipc}.

\subsubsection{Method} To understand to what extent these protestware caused supply chain disruptions, we conducted a qualitative retrospective observational study~\cite{mann2003observational} based on news articles and blogs online~\footnote{For brevity, we will refer to ``news articles and blogs'' simply as ``articles'' in this section, unless otherwise specified.}. This is because -- as protestware were launched in the past, only retrospective studies available data is feasible.
% It is known to be cheap and is often used in place of prospective studies (\eg interviews to gather new data) when they are considered unnecessary or ineffective~\cite{mann2003observational}. 
To start, we refer to articles to investigate effect of the protestware on the software supply chain.
Although our initial collection from~\cref{sec:collection} already contained news articles, this list could miss certain articles for a specific protestware. 
Thus, as a safety measure, we searched on the internet for the protestware itself. 
To ensure relevance to protestware, our search query is ``\texttt{"[protestware name]"~"protestware"}''~\footnote{The quotes around the keywords ensure that the search results include those exact terms.}.
We also snowballed~\cite{goodman1961snowball}, visiting any useful cited articles in the ones we already found.
Since we observed that these articles often cite a select few original articles, we did not look more than one page deep into the search results unless it was deemed necessary.
For instance, \texttt{node-ipc} is an impactful protestware yet we could not find quotes from the developer until the third page of results.
Since these articles may contain inaccurate information, we used our honest judgement and only considered the results if sufficient evidence is provided.
We conduct this study using articles over interviews because interviews pose similar threats, which we discuss further in~\cref{sec:threatsretrostudy} (threats to validity). 

\subsubsection{Results} 
We found a total of 90 articles with duplicates, resulting in 55 unique articles. Next, we present our qualitative findings based on different categories of protestware.
% \JC{Borrow reasoning from iot paper}
% \SR{Talk about how many of the protestware had a potential to affect supply chains and how many made it to the news and then talk about their impact.}
% \SR{We should also highlight that some of the protestware did not made into news, maybe because they did not do radical things? It would be nice to have some explanation about why these two not others.}
% \JC{Notes
% Awesome Prometheus Alerts - Russians cannot use - no news
% colors - ininite loop - no one can use - yes news
% faker - self-sabotage
% node-ipc - delete files for Russians - news
% Quake3e - game engine - Russians cannot use - no news
% SweetAlert2 - Russians cannot use - no news
% tasmota - Russians cannot use - no news
% left-pad self-sabotage - yes news
% }
% However, few articles reported on the protestwares' effects on the supply chain in sufficient detail. Much of the content are comments by others or technical information about the protestware. One news article briefly covered the protestware with 2 infinite-loops act like DoS's as well~\cite{fakerarchive,colorscommit} and estimated their effects to be in the range of thousands of developers~\cite{fakersnyk}, without providing any real evidence. 
% For instance, \texttt{faker.js} is used to create fake test data.

\textbf{Irreversible damage in targeted critical infrastructure.}
Vulnerable versions of \texttt{node-ipc} deleting files targeting Russian and Belarus users existed on NPM for less than 24 hours~\cite{nodeipcsnyk} which still reportedly affected large OSS projects. 
For instance, Vue.js~\cite{vuejs}, a popular JavaScript front-end framework with over 5 million weekly downloads~\cite{vuejsnpm}, always used the latest minor and patch versions of \texttt{node-ipc} instead of pinning a known safe version~\cite{nodeipcbleepingcomputer}. 
This inevitably caused Vue.js to use a vulnerable version, which reportedly affected its downstream users~\cite{vuejsnodeipcimpact}.
Furthermore, in response to \texttt{node-ipc}, Russian bank Sber advised their customer to stop updating their software due to concerns over malicious code~\cite{nodeipctheverge,nodeipcthevergerussian}.
%The developer of \texttt{node-ipc} also claimed to be swatted~\footnote{Swatting is an ``attack where someone finds a victim's address and alerts police to a fake emergency there''~\cite{nodeipcitpro}.}~\cite{nodeipcitpro}.
% No figures were reported regarding the number of users whose files deleted due to \texttt{node-ipc}, since this would require the user to know about \texttt{node-ipc} and actively report it.
% \JC{Do we want this last sentence.}

\textbf{Failures due to self-sabotage-based DoS.}
First, \texttt{left-pad} was one of the packages deleted in a trademark dispute over a package named ``kik''~\cite{leftpad}. 
NPM observed \myquote{hundreds of failures per minute, as dependent projects -- and their dependents, and their dependents... -- all failed when requesting the now-unpublished package}~\cite{leftpadnpmblog}. 
While another developer soon published his own functionally identical version of \texttt{left-pad}, errors continued because certain projects explicitly request version 0.0.3, whereas the new one was in 1.0.0. 
To solve this issue, they republished version 0.0.3 of the original \texttt{left-pad}.
The entire duration lasted 2.5 hours.
Regarding the restoration of \texttt{left-pad}, NPM CTO said \myquote{This action puts the wider interests of the community of NPM users at odds with the wishes of one author[developer]; we picked the needs of the many}~\cite{leftpad}. This highlights the impact a misbehaving trusted library can have on the entire community.

\textbf{Impact of other protestware.} 
We were able to find 24 articles covering \texttt{colors.js} and \texttt{faker.js} denying services, presumably because of their high potential for impact with millions of weekly downloads. 
For instance, Revenera, a software auditing company, reported that \textit{``82\% of audit service customers from Revenera in 2021 contained Node Module Packages. Of those, 94\% use colors.js while faker.js ranks at 67\%''}~\cite{revenera}. 
In another article, it is estimated that \texttt{colors.js} and \texttt{faker.js} impacted thousands of applications~\cite{colorsdarkreading}.
However, none of the articles reported specific numbers or any confirmed cases in terms of their impact. 
Futhermore, no articles were found for aforementioned 4/5 conditional DoS protestware~\cite{awesomeprometheusalertscommit,quake3e,sweetalertcommit, tasmotacommit}.
% We also found 24 articles for 6 non-destructive protestware~\cite{e2eakarev, esfiveextcommit, eventsourcecommit,nestjs-pino, peacenotwar,styledcomponentscode}, but no specific impacts were reported.
% \JC{Do we want to list citations or the protestware names? Citations saves space.}

\begin{SummaryBox}{Takeaway \cref{sec:supplychaineffects}: You can't avoid a deleted library...but you can choose not to always use the newest version}
Automatically updating software to the newest version is a common recommendation for end-users to ensure that their software has all known vulnerabilities fixed.
While using the newest version is generally a good idea, it may be catastrophic if one of the dependencies becomes a protestware.
\end{SummaryBox}

\subsection{Sentiment Analysis.}
% \JC{Since we use positive/negative, the word sentiment may make more sense.}
\label{sec:reactionanalysis}
% To measure the community reaction we qualitatively looked at the \textit{sentiments} in the comments posted under commits or issues that introduced protests. This implies that we naturally excluded the ones that did not have comments (\textcolor{red}{9} out of \textcolor{red}{21}).  
\subsubsection{Method.} 
During this investigation, we used the following codes to label the general sentiments for a given commit: positive, negative, and neutral. This coding was performed asynchronously by two authors aggregating the sentiments of the comments and reactions emojis on comments as prominent indicators for a given commit into a single label. The label was then discussed between both authors until a consensus was reached on 2 initial disagreements out of 19 labels. During this analysis, we also noted if a given protestware is still active. To determine if the protestware was currently active, the current version of the repository was checked for the pieces of the code determined to be inducing the protest. It is essential to note that in qualitative studies, in a given context, any mention of statistics or counts for specific codes only holds for that context. 
% \JC{\cite{miller2022did} has 5 categories of toxic comments on GitHub in \$4, maybe we can borrow them? This is really only useful if the number of comments in our corpus is large, otherwise positive/negative is sufficient.}

% \SR{Add the disclaimer that gross count might not show a comprehensive picture.}

\subsubsection{Results.} In this section, we present the results of our sentiment analysis. In total, we found that 8/21 altered protestware (with available commit links) contained negative, 3 contained more positive, 1 contained neutral sentiments, and the remaining having no comments associated the commit. We also found that 11/21 are still active. 

\textbf{Positive sentiments and supports.}  
Some protestware (3 total marked) received positive comments from the community supporting political messages in the software. However, the engagement was considerably lower than the ones with negative comments, except for \texttt{color.js}, which contained a variety of positive, negative, and off topic comments. In one instance (\texttt{Evolution}), there were only two positive reaction emojis on the commit that changed the background login image to something political. Another instance in \texttt{colors.js} is one user giving a supporting message for developers within a long line of comments and 16 positive reactions to the comment saying the following quote:

\myquote{Bless all these people who've been maintaining small but very important things for this long [Thumbs Up Emoji]}~\cite{colorscommit}.

There were some comments anticipating that the commit will receive a lot of attention from media sources (news articles, blogs, etc) so that many people would view it, saying something like:
\myquote{Hi mom! I'm on TV!}~\cite{colorscommit}.
The final form of positive sentiment seen is through American patriotism. The user wishes to express a positive emotion for being in America and the pride of associating that country. The comments of this nature tend to be short and repetitive:
\myquote{America Babyyy [4x American Flag Emojis]} and 
\myquote{MakeAmericaGreatAgain}~\cite{colorscommit}, which was former president Donald Trump's campaign slogan. We note that 1 (out of 3) were reverted back to normal even with the positive support. The primary reason was the importance of the package and the behavior in them. The one reverted was left-pad which broke enough projects to get media attention. left-pad~\cite{leftpad} was reinitialized by NPM because the CTO of NPM felt that the needs of their community outweighed the actions of one developer. We also note that all three of them were universally affecting their downstream users and were non-transparent too.

% \SR{@Tanner, the following takeaway is weak and confusing. Can we rewrite it -- specifically by highlighting that ``support'' is more correlated with the potential of media attention attracting like-minded people?}
\begin{SummaryBox}{Takeaway \cref{sec:reactionanalysis} (1/2): Positive sentiments are correlated with political messages}
Positive support from the user community (with like-minded individuals) exists for a small number of protestware -- which is mostly correlated with their political stance. 
\end{SummaryBox}

\textbf{Negative sentiments and pushbacks.}
Many protestware (8 marked) received pushback with negative comments surrounding the insertion of protestware. These protestware were all surrounding the Russo-Ukraine war and generally (7 out of 8) targeted users in Russia. The typical behaviors seen in these comments include asking the developer to revert the commit, direct negative opinions/insults, saying it does not help anything related to the source of protest, negative reaction emojis to previously positive comments, and some with an understanding of reasoning, but disagreeing with the implementation. Examples of asking to revert back include multiple users asking the developers not to mix political ideologies in their code: \myquote{Stop politics}, \myquote{Just stop messing politics and javascript.}~\cite{esfiveextcommit}. 
Some other negative opinions are:
\myquote{all ur message looks like propaganda for stupid peoples}~\cite{esfiveextcommit} and 
\myquote{what a stupid code here!}~\cite{eventsourcecommit}.
Others describe that the protestware will not help in the provided conflict or rhetorically give affirmation:
\myquote{How exacly this sh$^*$t must help?
This is war, idiots.}~\cite{node-ipc} and
\myquote{Of course Putin is using JS, he will certainly see your message.}~\cite{esfiveextcommit}. 
Finally, we saw comments acknowledging the developer's sentiment but still requesting the removal of the protestware in \texttt{quake3e}: \myquote{This sort of action is deeply disappointing, I hope you reconsider on this. Holding ordinary enthusiasts to account for the actions of their government will not achieve anything useful for anyone}~\cite{quake3e}.

We observe that only 3 were transparent protestware, indicating their intentions in the README file as well. We also note that 3 (out of 8) of the instances reverted their software back to normal while the other five have protestware still in their software (as of May 2024). So even though the user comments on protestware indicate a stronger overall dislike of the insertion, the developers only listened a little less than half the time.
% The impact here shows that open source developers may take into account the opinions of their user base, but can also largely deviate from that path. 

\begin{SummaryBox}{Takeaway \cref{sec:reactionanalysis} (2/2): Pushback does not imply changes in developers actions}
Even with negative comments from the OSS community, a developer can still maintain their own beliefs integrated into their code. In fact, only 3 (out of 8) cases, developers reverted their code.
\end{SummaryBox}
% \JC{Do we have an exact count, how many comments are positive and how many are negative? Can I just count the examples to get the answer here or no?} \TF{No, we didn't track individual comments, just the general 'vibe' of each}

% This could indicate that even through negative comments developers can still stick to their beliefs. 

% Some comments were a little more understanding saying they do not like the war either, but protestware was not the solution. There was only one repository with only positive comments which was 'Evolution' and had 2 positive comments about changing the background login image from something neutral to an image with aggressive language against Russia as shown in Figure ~\ref{fig:changeImage}. Although most of the repositories saw negative comments, their popularity statistics did not decrease afterward. 

\subsection{Usage Trends} 
\label{sec:popularityanalysis}
To understand how the protestware affected the trust dynamics in the OSS community, we conducted a usage trend analysis. 
We limited the scope of this analysis to only libraries since the ``trust dynamics'' are meaningful in the context of supply chain dependency. 

\textbf{Method.} We conducted a comparative study of \textit{usage} counts (number of dependents and weekly download counts) of a given library from when they were converted into protestware to the present time.
% \SR{Talk about why we are using download count even though it can vary drastically.}
% To understand the usage trends of the libraries that turned into protestware, we conducted a comparative analysis of their download counts from the point they were turned into protestware vs. the current time. 
%Watches provides the count of users tracking the project, while star count indicates the popularity of a project in the form of social proof and appreciation and forks refer to copies of a project that users can create to make their own independent versions of it. 
% These stats can be a good indicator of the impact of the protestware that alters README as their goal is to reach the OSS community.
% Although they are not a perfect proxy for real-world deployment counts, these stats still have moderate correlation to real-world usage~\cite{kochfault}.
% These stats are not trivial.
% For instance, a study on 700+ developers found that 73\% of them consider the number of GitHub stars before using the OSS, indicating its importance~\cite{borges2018s}.
% \JC{Added above.}
Since 12/\totalprotestwarecount
of the libraries that were turned into protestware are written in JavaScript, we used the number of \textit{dependents} and weekly download counts from the Node Package Manager (NPM)\footnote{NPM is widely used to host Node.Js libraries~\cite{npm}.} to measure the usage trends\footnote{PHP also has some metrics for usage, however since the metrics are not directly comparable and both our libraries(\texttt{yandex} and \texttt{Evolution}) have extremely low dependents and no valid snapshots in the year of the commit, we decided not to include them.}. Dependent counts indicate how many libraries in the NPM repository are directly or indirectly depended on a given library, \ie the downstream software in the supply chain. Weekly download counts indicate how many times a given library was downloaded from the repository. Although the number of weekly downloads can fluctuate drastically from week to week, however, maintaining a high weekly count would indicate that the trustworthiness is unaffected.
To understand how the usage changed, we compare the counts from before~\footnote{When there is no date that's both close and before, we use one that's slightly after.} a software turned into protestware to the current count. 
To find the counts from the past, we used the WayBack machine~\cite{Wayback}. Although the gross count across a two-year time span does not show a comprehensive view, the overall pattern can be seen. % \JC{``comprehensive view'' and ``overall pattern'' is vague and can be confusing. Are we saying the two-year time span doesn't show everything but we can probably extrapolate?}\TF{I think we meant that it's not very fine-grained (only a few data points), but we can roughly interpolate(get an estimate) between the two values and see if they were increasing or decreasing }. 
% Note that each of these measures are not completely accurate with regard to actual developer usage and user impact. 
% However, each provide a fairly good estimate of the popularity within the repository. 
% \JC{Rewrite this part to properly reflect initial intentions. What were they again?}
% \TF{I think most of it sounds good, but I think we wanted to clarify the intentions aren't saying they are perfect measures (I added some at the end). You probably should rephrase it and potentially add the source for the last claim. }
% \JC{Noted. Done above}

% \textbf{Results -- Community push-backed, but popularity increased.} %We never said community pushed back yet, maybe we want to flip the sections (Have reaction analysis first)
\textbf{Results -- General dependency increased.} 
In total, we analyzed a collection of 12 protestwere  JavaScript libraries. 
In this collection, 7 are targeting specific users,  5 are publicized, and 5 are still active protestware. 
Our analysis revealed that 10 out of 12 libraries have an increased number of dependency counts (\cref{tab:depStats}). Note that for \texttt{left-pad} a protestware from \textit{halted service} category, we were unable to determine the dependency count for a time near the removal using the Wayback machine. After NPM reinstated the package, it had 1.4 million weekly downloads, although it is marked as deprecated~\cite{leftpadrepo}. 
The weekly download count does not reveal any noticable patterns overall. 
The library with the largest sizable decrease in weekly download count was \texttt{EventSource}, going from 8.6 million to 4.4 million downloads each week
% \JC{Nice to have citations to the waybackmachine links. Not just here but everywhere}. 
%Another library, 'left-pad', went from 439K to 1.4 million\JC{The 1.4 million version is published by another dev, see~\cref{sec:supplychaineffects}. It is also now deprecated. \url{https://www.npmjs.com/package/left-pad}}, although it was on a larger time scale since the left-pad changes occurred in 2016. 
Overall, an increase in dependency count for protestware with the potential of affecting the software supply chain reveals a surprising aspect of the trust dynamics of the OSS community. A close look at the top-7 most popular NPM libraries with millions of current downloads (i.e., \texttt{es5-ext}, \texttt{EventSource}, \texttt{colors.js}, \texttt{faker.js}, \texttt{styled-components}, \texttt{pnpm}, and \texttt{left-pad}) indicates that 5 out of 7 
% \JC{Which 5? Is this indicated anywhere.} \TF{The ones not described as active}
of them currently do not contain protestware, where \texttt{es5-ext} and \texttt{color.js} still remain active. 

% \JC{Does this takeaway apply to all or a subset of protestware?}
% \TF{Yeah, not sure if this takeaway applies because we haven't talked about the sentiment yet, but as far as the claim goes: it holds for es5-ext with negative sentiment, but not colors.js which we marked with positive sentiment.}
% \JC{We can do a combined takeaway in the latter section, currently it's sentiment analysis.}
% \JC{What if the order is (1) effects on supply chain, (2) sentiment, (3) github/npm stats? I think effects on supply is what most would care about. Sentiment then github/npm stats means that we can incoorporate sentiment into the takeway.}

\begin{SummaryBox}{Takeaway \cref{sec:popularityanalysis}: Increase in protestware library dependencies -- misplaced trust?}
Even with the potential to affect the software supply chain, the dependency count of 10 out of 12 protestware increased after turning them into protestware. Some protestware still remains active with millions of weekly downloads.
\end{SummaryBox}

%Old Version
% \begin{table*}[htbp] % Table for comparing dependencies
% \centering
% \caption{Dependents Comparison Table of JavaScript Libraries Dependents Count and Downloads}
% \begin{tabularx}{\textwidth}{|l|c|c|c|c|c|c|c|}
% \hline
% \textbf{name} & \textbf{old dependents} & \textbf{current dependents} & \textbf{dependents \% diff} & \textbf{old weekly downloads} & \textbf{current weekly downloads} & \textbf{downloads \% diff} \\
% \hline
% peacenotwar & 1 & 2 & +1 & 483 & 1342 & +20 \% \\
% \hline
% node-ipc & 355 & 398 & -2 & 892523 & 463895& -20 \% \\
% \hline
% es5-ext & 216 & 301 & +2 & 12547294 & 9426242
% & +16.67 \% \\
% \hline
% EventSource & 560 & 787 & +2 & 8666668 & 4443843 & +25 \% \\
% \hline
% SweetAlert2 & 901 & 1844 & 0 & 432118 & 578017 & 0 \% \\
% \hline
% colors.js & 18958 & 22111 & +2 & 22417827 & 15874535 & +11.11 \% \\
% \hline
% faker.js & 2570 & 2632 & -1 & 1712938 & 1586276 & -6.25 \% \\
% \hline
% e2eakarev & 0 & 0 & +2 & 2 & 1 & +22.22 \% \\
% \hline
% styled-components & 17095 & 24414 & 0 & 3907380 & 5395745 & 0 \% \\
% \hline
% left-pad & -- & 534 & 33 & 439841 & 1436312 & -50 \% \\
% \hline
% \end{tabularx}
% \end{table*}
%GPT calculation
% \newcommand{\posdiff}[1]{\textcolor{green!50!black}{#1}}

\begin{table}[htbp] % Table for comparing dependencies
% \JC{We need citations to the links showing the download counts. Just showing the two general websites seems lazy.}
% \SR{Let's keep a note for it, low priority for now.}

\scriptsize
\centering
\caption{Dependents Comparison Table of JavaScript Libraries Dependents Count and Downloads. Data collected using ~\cite{Wayback} and from ~\cite{npm}. ($^*$ - the library no longer exists, so the most recent snapshot was used)
% \JC{Looking at the rightmost column, it doesn't seem like ``general popularity increased''? }
}
\label{tab:depStats}
\begin{tabular}{|l|r|r|r||r|r|r|r|}
\hline
\textbf{name} & \textbf{old dep.} & \textbf{curr. dep.} & \textbf{dep. \% diff} & \textbf{old weekly downloads} & \textbf{curr. weekly downloads} & \textbf{downloads \% diff} \\
\hline
peacenotwar & 1 & 2 & \posdiff{+100 \%} & 483 & 1342$^*$ & \posdiff{+177.8 \%} \\
\hline
node-ipc & 355 & 398 & \posdiff{+12.1 \%} & 892523 & 463895 & \textcolor{red}{-48.0 \%} \\
\hline
es5-ext & 216 & 301 & \posdiff{+39.4 \%} & 12547294 & 9426242 & \textcolor{red}{-24.8 \%} \\
\hline
EventSource & 560 & 787 & \posdiff{+40.5 \%} & 8666668 & 4443843 & \textcolor{red}{-48.7 \%} \\
\hline
SweetAlert2 & 901 & 1844 & \posdiff{+104.7 \%} & 432118 & 578017 & \posdiff{+33.8 \%} \\
\hline
colors.js & 18958 & 22111 & \posdiff{+16.7 \%} & 22417827 & 15874535 & \textcolor{red}{-29.2 \%} \\
\hline
faker.js & 2570 & 2632 & \posdiff{+2.4 \%} & 1712938 & 1586276 & \textcolor{red}{-7.4 \%} \\
\hline
e2eakarev & 0 & 0 & N/A & 2 & 1 & \textcolor{red}{-50.0 \%} \\
\hline
styled-components & 17095 & 24414 & \posdiff{+42.8 \%} & 3907380 & 5395745 & \posdiff{+38.1 \%} \\
\hline
pnpm & 22 & 207 & \posdiff{+840.9 \%} & 632655 & 15403564 & \posdiff{+2334.7 \%} \\
\hline
nestjs-pino & 35 & 177 & \posdiff{+405.7 \%} & 68901 & 463038 & \posdiff{+572.0 \%} \\
\hline
left-pad & -- & 534 & N/A & 439841 & 1436312 & \posdiff{+226.6 \%} \\
\hline
\end{tabular}
\vspace{-1em}
\end{table}

\section{Discussion}

% \subsection{Protestware Developers' Stance}
% \JC{``Nowak said using his open source library for activism did not affect his credibility among the wider community, but he did receive a handful of angry responses at the beginning.'' Is this any use? \cite{protestwareontherise}. Not really a stance.}
% \JC{``Miller has defended his peacenotwar module on GitHub, saying "this is all public, documented, licensed and open source." '' ~\cite{peacenotwarregister}. Any use? We kind of have it in }

\subsection{Ethical Considerations}

Following Kula \etal~\cite{kula2022in}, we also considered labeling protestware as malignant or benign based on how a protestware behaves. However, labeling a protestware as benign or malicious might induce \textit{i)} political bias towards the protestware community, and \textit{ii)} can easily be mistaken for our political stances. 
Thus, we chose to remove this label from our set of characteristics. 
Instead, we capture all of the interesting features for an objective stance can be made. 
Furthermore, all disclosed information in this work are publicly available.

\subsection{Threats to Validity}
\subsubsection{Construct Validity}
While prior work has defined protestware to be a form of supply chain attack by altering an OSS~\cite{Protestware}, we extend the definition to include anything related to software and protests. We believe this approach gives readers a comprehensive perspective, providing additional assurance to the validity of the work. In~\textbf{RQ1}--\cref{sec:char}, we operationalized the constructs of ``nature of inducing protests'', ``nature of targeting users'', and ``nature of transparency''.
However, we acknowledge that there may other constructs that were not considered. In~\textbf{RQ2}--\cref{sec:aftermath}, we operationalized the constructs of ``effects on the supply chain'', ``sentiment'', and ``usage trends''. 
While sentiment~\cite{taboada2016sentiment,medhat2014sentiment, wankhade2022survey, hussein2018survey} and usage trends~\cite{dey2018software, zerouali2019diversity} are commonly used metrics in prior works, we used a loose definition for ``effects on the supply chain'', as information was generally limited.
% Note that much of the work in sentiment analysis involves automation, but this was not needed or the focus of our scale of study.
% TF: Not sure what you mean by automation?
\subsubsection{Internal Validity}

\textbf{Searching for Protestware.}
We relied on external sources to curate our list of protestware, which ultimately limits our dataset to those protestware already known in the wild.
This particularly concerns the ones that are altered software because these are the ones that may have an incentive to be secretive.
%Although we rigorously reviewed each candidate's protestware presented in said sources to ensure that it's a valid protestware, there is no guarantee that these sources are comprehensive.
%For instance, some of these protestware samples are detected via the commit message~\cite{evoCMScommit}.
% It is possible a commit message does not hint that the change is for protesting.
However, we believe it is unlikely that we missed a protestware, especially impactful ones, because it went unnoticed and unreported by users. This conjecture is supported by the observation that people are often vocal about their experiences and observations in the age of the Internet and social media~\cite{cinelli2021echo}.
\textbf{Recency Bias in Dataset.}
One limitation of our study is that our set of protestware are all from 2013~\cite{freemyinternet} or later. 
The set of protestware that altered software is particularly biased towards recent years, with none being made before 2022. 
% This bias is largely due to using Internet search for protestware articles and lists, which may be biased toward recent articles.
% Searching with ``protestware after:2000-01-01 before:2010-12-31'' on Google, however, returned only 7 results.
% While ``protestware after:2011-01-01 before:2020-12-31'' returned more results, no articles on the first page seemed to provide protestware samples not already in our corpus, if any at all.
We believe that the lack of results prior to 2020 and particularly before 2010 may be because protestware is a fairly new term and perhaps even concept.
No concrete proof can be made on when the term protestware is used, but there exists evidence to suggest that it is recent. 
For instance, the term ``protestware'' is known to be popularly used in 2022 after the \texttt{node-ipc}~\cite{puppylinux} incident~\cite{protestwareisontherise}.
For these reasons, we believe our dataset is reflective of the truth.
\textbf{Qualitative Coding.}
Our study is largely, though not entirely, qualitative. 
Such studies necessitate subjective evaluations which will induce human bias into all manual codings~\cite{guest2012applied,miles1994qualitative}.
For instance, our set of characteristics in~\cref{sec:char} may not be exhaustive; it is entirely possible, though we conjecture this to be unlikely, that another researcher can extend our set.
In another example, various researchers may interpret positive or negative sentiment in~\cref{sec:aftermath} differently based on personal experiences.
However, to mitigate said bias and increase trustworthiness of each qualitative stage of our study, we had multiple authors rigorously validate each other's codes.
A potential alternative may be the use of large language models, but they may introduce their own biases, even with fine-tuning~\cite{gallegos2023bias, herrera2023large, navigli2023biases}. Lastly, because it is qualitative in nature, our study might not generalize to the data samples outside of our corpus.
\textbf{Retrospective Study via News Articles and Blogs.}
\label{sec:threatsretrostudy}
For the supply chain impact analysis, we used news articles and blogs, which may contain biases, dramatization, and selectively cover events~\cite{mullainathan2002media}.
However, given the cost-benefit tradeoff, we believe using these data sources is reasonable because the alternative, for instance, is running an internet-wide survey without any guarantee of a better outcome. The fact that most of the incidents in our case happened in the past, the user study-based approach might suffer from a similar bias. To lower the probability of false facts, we cross-checked the facts claimed in the news across multiple sources. We take into consideration the reputation of the publisher by checking if it's in the Iffy list, which contains a list of unreliable sources~\cite{iffy}.
The Iffy list has been used in existing literature~\cite{hanley2024specious} and other works have also used news articles as a part of their study~\cite{anandayuvaraj2022reflecting,chen2024contents}.

\subsubsection{External Validity}
For this study, we do not make any claims on generalizability. 
Rather, our focus is on analyzing protestware that is readily available for study. 
Thus, we see no threats to external validity.
\section{Related Work}

Our work studies protestware, which can be a threat to the software supply chain. We first look at works directly related to protestware, and then we present other works in the general realm of software supply chain.

\subsection{Protestware and Malware}
To the best of our knowledge, no prior work has systematically and comprehensively studied protestware in the wild. We know of only two works specifically focus on protestware~\cite{kula2022in, Protestware} and another work that includes protestware as a minor part~\cite{wermke2023always}. 
Kula~\etal proposes 3 categories of protestware by giving a few examples of each without systematically collecting a comprehensive set of protestware~\cite{kula2022in}. Cheong~\etal proposed ethical guidelines for OSS developers potentially turning their library into protestware.
As part of a larger interview with 25 developers, Wermke~\etal surveyed them for their views on various aspects the open-source supply chain, including protestware (\ie \texttt{node-ipc})~\cite{wermke2023always}.
Some malware studies focus on a in-depth investigation of a single family. For instance, Antonakakis~\etal studied the Mirai botnet~\cite{antonakakis2017understanding}, while Herwig~\etal studied the Hajime botnet~\cite{herwig2019measurement}.
However, they can also be investigated comprehensively like our study of protestware. 
For example, Cozzi~\etal conducted the first comprehensive study, collecting and analyzing 10K malware samples~\cite{cozzi2018understanding}.
Alrawi~\etal expanded on this dataset and investigated the lifecycle of 166K Linux-based IoT malware~\cite{alrawi2021circle}.

\subsection{Packages and their Managers} 
One strand of work involves conducting a retrospective study to investigate packages and their managers, \eg PyPI, npm, RubyGems.
For instance, Zimmermann~\etal studied dependency relationships, project maintainers, and known security issues for npm packages~\cite{zimmermann2019small}.
Zahan~\etal analyzed the metadata of npm packages for signals of weak security~\cite{zahan2022weak}.
Valiev~\etal investigated the factors affecting the sustainability of the PyPI ecosystem~\cite{valiev2018ecosystem}.
Gonzalez~\etal created a tool to automaticaly detect malicious packages using only GitHub commits \cite{gonzalez2021anomalicious}.
Duan~\etal studied over one million packages from PyPI, npm, and RubyGems, finding 339 malicious packages~\cite{duan2020towards}.

\subsection{Human Factors}
Another facet of work is to understand the human factors in the software supply chain. 
For example, Abdalkareem~\etal surveyed 88 Node.js developers to assess their opinions on the benefits and drawbacks of using trivial packages~\cite{abdalkareem2017developers}.
Wermke~\etal interviewed 25 developers to understand their project processes, decisions, and considerations regarding open-source software~\cite{wermke2023always}.
Miller~\etal interviewed 33 developers to study how they manage open-source dependency abandonment, realizing that the developers are often left with minimal support or guidance~\cite{miller2023we}.
In another study, Miller~\etal investigated online toxicity in the discussions of open-source software forums, finding entitled,
demanding, arrogant, and insulting comments~\cite{miller2022did}.
Bogart~\etal studied how developers and organizations handle changes in dependencies through a series of interviews, discovering many challenges and headwinds~\cite{bogart2015breaks,bogart2016break}.

\section{Conclusion}
\label{sec:conclusion}

In this work, we curated the first comprehensive dataset containing \totalprotestwarecount protestware.
We then investigated the protestware through a systematic and iterative process to (1) identify useful characteristics of protestware and (2) understand the impacts of disruptive protestware on the supply chain, community sentiment, and usage trends. 
Three characteristics were identified: (i) the nature of inducing protests, (ii) the nature of affecting users, and (iii) the nature of transparency.
A taxonomy was created to describe the different ways of inducing protests.
Our aftermath analysis showed that disruptive protestware, namely \texttt{left-pad} and \texttt{node-ipc}, can have profound negative impact on downstream users.
In addition, usage generally increases even after inserting protestware in code.
Furthermore, we found that developers maintain their own beliefs even with community pushback.
The implication for regular developers is that they can never fully trust OSS, and they should have a contingency plan if and when their software fails due to an abuse of their upstream supply chain.
Future work could investigate ways to automatically detect protestware so that users are notified immediately rather than retroactively.
% Informed by this study, in the future, it will be interesting to investigate this trust dynamics further.

\section{Data Availability}
A replication package is uploaded using Zenodo~\cite{suppmaterial}. The package contains spreadsheets with protestware, coded labels and themes, collected articles, and the appendix.

\newpage
\bibliographystyle{ACM-Reference-Format}
\bibliography{references}

\newpage 
\section*{Appendix}

\begin{figure}[h]
    \centering
    \scriptsize 
    \begin{lstlisting}[style=base, 
        language=html, 
        label={colorsflag},        caption={Example of ideology promotion in american.js of the colors project~\cite{colorscommit}.}]
module.exports = function americanFlag () {
console.log('LIBERTY LIBERTY LIBERTY'.yellow);
console.log('LIBERTY LIBERTY LIBERTY'.america);
console.log('LIBERTY LIBERTY LIBERTY'.yellow);
let flag = "\
\
                             !\
         H|H|H|H|H           H__________________________________\
         H|$|$|$|H           H|* * * * * *|---------------------|\
         H|$|8|$|H           H| * * * * * |---------------------|\
         H|$|$|$|H           H|* * * * * *|---------------------|\
         H|H|H|H|H           H| * * * * * |---------------------|\
         H|H|H|H|H           H|---------------------------------|\
      ===============        H|---------------------------------|\
        /| _   _ |\          H|---------------------------------|\
        (| O   O |)          H|---------------------------------|\
        /|   U   |\          H-----------------------------------\
         |  \=/  |           H\
          \_..._/            H\
          _|\I/|_            H\
  _______/\| H |/\_______    H\
 /       \ \   / /       \   H\
|         \ | | /         |  H\
|          ||o||          |  H\
|    |     ||o||     |    |  H\
|    |     ||o||     |    |  H   Carl Pilcher\
";

console.log(flag);

}
    \end{lstlisting}
\end{figure}

\begin{figure}[h]
\small
\begin{lstlisting}[style=base,
language=,
caption={Added protest message in README.md in pnpm~\cite{pnpmchange}.}, 
label={alteredDoc}]
> # UKRAINE NEEDS YOUR HELP NOW!
>
> I'm the creator of this project and I'm Ukrainian.
>
> **My country, Ukraine, [is being invaded by the Russian Federation, right now](https://www.bbc.com/news/...)**. I've fled Kyiv and now I'm safe with my family in the western part of Ukraine. At least for now.
> Russia is hitting target all over my country by ballistic missiles.
>
> **Please, save me and help to save my country!**
>
> Ukrainian National Bank opened [an account to Raise Funds for Ukraine's Armed Forces](https://bank.gov.ua/en/news/all...):
...
> You can also donate to [charity supporting Ukrainian army](https://savelife.in.ua/en/donate/).
>
> **THANK YOU!**
\end{lstlisting}
\end{figure}

\end{document}